\documentclass[12pt,twocolumn]{emulateapj}
\pdfoutput=1 
\usepackage{slashbox}
\usepackage{diagbox}
 \usepackage{ulem}
\usepackage[table]{xcolor}

\usepackage[page]{appendix}

\begin{document}

\submitted{Submitted to Astrophysical Journal June 24, 2015; Accepted April 4, 2016}

\title{Morphology and Molecular Gas Fractions of Local Luminous Infrared Galaxies as a Function of Infrared Luminosity and Merger Stage.}
\author{
K. L. Larson\altaffilmark{1,2}, 
D. B. Sanders\altaffilmark{1,3}, 
J. E.  Barnes\altaffilmark{1,4}, 
C. M. Ishida\altaffilmark{1,5}, 
A. S. Evans\altaffilmark{6,7,8}, 
V. U\altaffilmark{9},
J. M. Mazzarella\altaffilmark{2},
D.-C. Kim\altaffilmark{7}, 
G. C. Privon\altaffilmark{10,11},
I. F. Mirabel\altaffilmark{12,13},
H. A. Flewelling\altaffilmark{1}
}

\altaffiltext{1}{University of Hawaii, Institute for Astronomy, 2680 Woodlawn Dr., Honolulu, HI 96822, USA; klarson@ifa.hawaii.edu}
\altaffiltext{2}{Infrared Processing and Analysis Center, California Institute of Technology, 1200 E. California Blvd., Pasadena, CA 91125, USA}
\altaffiltext{3}{Visiting Astronomer, Research School of Astronomy and Astrophysics, Australian National University, Cotter Road, Weston Creek, ACT 2611, Australia} 
\altaffiltext{4}{Visiting Professor, Yukawa Institute for Theoretical Physics, Kyoto University, Kitashirakawa Oiwakecho, Sakyo-ku, Kyoto 606-8502 Japan} 
\altaffiltext{5}{Department of Physics and Astronomy, University of Hawaii at Hilo, Hilo, HI 96720-4091, USA}
\altaffiltext{6}{Department of Astronomy, University of Virginia, P.O. Box 400325, Charlottesville, VA  22904, USA}
\altaffiltext{7}{National Radio Astronomy Observatory, Edgemont Road, Charlottesville, VA 22904, USA}
\altaffiltext{8}{Visiting Astronomer, Infrared Processing and Analysis Center, California Institute of Technology, 1200 E. California Blvd., Pasadena, CA 91125, USA}
\altaffiltext{9}{Department of Physics and Astronomy, University of California, Riverside, 900 University Avenue, Riverside, CA  92521, USA}
\altaffiltext{10}{Departamento de Astronom\'ia, Universidad de Concepci\'on, Casilla 160-C, Concepci\'on, Chile}
\altaffiltext{11}{Instituto de Astrof\'isica, Facultad de F\'isica, Pontificia Universidad Cat\'olica de Chile, 306, Santiago 22, Chile}
\altaffiltext{12}{Instituto de Astronom\'ia y F\'isica del Espacio. CONICET. Universidad de Buenos Aires. Ciudad Universitaria.1428 Buenos Aires, Argentina} 
\altaffiltext{13}{Laboratoire AIM, CEA/DSM/Irfu/Service d'Astrophysique, Centre de Saclay, B\^{a}t. 709, FR-91191 Gif-sur-Yvette Cedex, France}
%
\begin{abstract}

We present a new, detailed analysis of the morphologies and molecular gas fractions for a complete sample of 65 local luminous infrared galaxies (LIRGs) from the Great Observatories All-Sky LIRG Survey (GOALS) using high resolution $I$-band images from {\it The Hubble Space Telescope}, the University of Hawaii 2.2m Telescope and the Pan-STARRS1 Survey.  Our classification scheme includes single undisturbed galaxies, minor mergers, and major mergers, with the latter divided into five distinct stages from pre-first pericenter passage to final nuclear coalescence.  We find that major mergers of molecular gas-rich spirals clearly play a major role for all sources with $L_{\rm IR} > 10^{11.5} L_\odot $; however, below this luminosity threshold, minor mergers and secular processes dominate.  Additionally,  galaxies do not reach   $L_{\rm IR} > 10^{12.0} L_\odot $ until late in the merger process when both disks are near final coalescence.  The mean molecular gas fraction (MGF $= M_{\rm H_2} / (M_* + M_{\rm H_2})$) for non-interacting and early-stage major merger LIRGs is 18$\pm 2$\%,  which increases to 33$\pm 3$\%, for intermediate stage major merger LIRGs, consistent with the hypothesis that, during the early-mid stages of major mergers, most of the initial large reservoir of atomic gas (HI) at large galactocentric radii is swept inward where it is converted into molecular gas (H$_2$).   
\end{abstract}


\section{Introduction}

Luminous infrared galaxies (LIRGs, $L_{\rm IR}(8-1000\mu m) > 10^{11} L_\odot$) are galaxies where intense infrared emission is fueled by star formation and active galactic nuclei (AGN) \citep[e.g.][]{Rieke:1972aa, Joseph:1985aa, Soifer:1987aa, Sanders:1988aa}. 
Although relatively rare in the local universe ($z < 0.3$), their number density still exceeds that of optically selected starburst and Seyfert galaxies at comparable redshift and bolometric luminosity \citep[e.g.][]{Soifer:1987aa}.  For the more extreme ultra-luminous objects (ULIRGs: $L_{\rm IR} (8-1000\mu m) > 10^{12} L_\odot$), whose infrared luminosity is equal to the bolometric luminosity of optically selected quasi-stellar objects (QSOs), the number density remains a factor of $\sim 1.5 - 3\times$ larger than that for QSOs at comparable bolometric luminosity \citep[e.g.,][]{Sanders:2003aa}.

Deep infrared surveys have shown that the number density of (U)LIRGs increases rapidly with increasing redshift, e.g. $\Phi (z) \propto (1+z)^{3-5}$, where the exponent increases with increasing $L_{\rm IR}$ \citep{Kim:1998ab}.   At $z > 1$, the bolometric infrared luminosity of the population of LIRGs exceeds that of the total UV-optical luminosity output of all galaxies at a given redshift, while the relative contribution from ULIRGs increases to where they equal or exceed that from LIRGs \citep[e.g.,][]{LeFloch:2005aa}.    What powers (U)LIRGs, as well as understanding their relationship to galaxy evolution in general, continues to be the subject of intense research and debate \citep[e.g.][]{Joseph:1999aa, Sanders:1999aa}. While a degree of consensus has been reached in understanding the origin and evolution of (U)LIRGs in the local universe, there continue to be conflicting views of the origin and evolution of (U)LIRGs at higher redshift \citep[e.g.][]{Elbaz:2007aa, Tacconi:2008aa, Hung:2013aa}, where the surface brightness dimming and decreased spatial resolution are the major complicating factors in interpreting the multi-wavelength properties of individual sources. Deeper, and higher resolution data will clearly enhance our ability to understand the nature of (U)LIRGs at high redshift, but it is also important to continue studies of nearby sources in order to have a well understood sample for comparison with their high-redshift counterparts. 

According to the hierarchical formation model of galaxies, galaxies build up mass over time through interactions and mergers \citep{White:1978aa, Barnes:1992aa}.  Disk galaxy mergers create gravitational torques that cause gaseous dissipation and inflows in the galaxies \citep{Barnes:1996aa,Mihos:1996aa}. Gaseous inflows then induce star formation and can feed powerful AGN. 
To truly understand the processes involved in galaxy evolution we need to trace how galaxy properties change with galaxy morphology over time.

A useful sample for studying the properties of local (U)LIRGs is the Great Observatories All-Sky LIRG Survey \citep[GOALS:][]{Armus:2009aa}, a flux-limited sample ($S_{60\mu m} > 5.24$\,Jy) of 203 galaxies with $L_{\rm IR} > 10^{11} L_\odot$ selected from the $IRAS$ Revised Bright Galaxy Survey \citep[RBGS:][]{Sanders:2003aa}.   Extensive multi-wavelength (radio to X-ray) imaging and spectroscopic data have been obtained for different subsamples of the GOALS\footnote{(http://goals.ipac.caltech.edu)} sources, with the most extensive coverage being that for the sample of 65 GOALS objects discussed in this paper.   
Observations have shown that all ULIRGs and many LIRGS in the local universe involve strong tidal interactions and mergers between molecular gas-rich disk galaxies \citep[e.g.,][]{Haan:2011aa}. These galaxies may represent an evolutionary stage in the formation of quasars and massive ellipticals from gas-rich mergers \citep{Sanders:1988aa, Genzel:2001aa}.

Detailed morphology study of LIRGs is important to determine the role of interactions and mergers in the evolution of infrared luminosities and molecular gas fractions. Our previous studies of LIRG morphology either focused on only the most luminous sources \citep[e.g.,][]{Kim:2013aa, Kim:1998aa, Sanders:1988aa} or relied on lower resolution ($\sim 2 ''$) data \citep[e.g.,][]{Stierwalt:2013aa}. Here we use higher resolution $Hubble$ $Space$ $Telescope$ (HST), The University of Hawaii (UH) 2.2\,m and Pan-STARRS1 (PS1) optical data to determine the galaxy merger stage across the full range on LIRGs ($10^{11} L_\odot <  L_{\rm IR} < 10^{12.5} L_\odot $).

The total mass of molecular gas represents the fuel immediately available for forming new stars which are presumably responsible for powering the large observed infrared luminosity. 
The galaxy molecular gas fraction (MGF), or the ratio of total mass of molecular gas to the total mass (gas mass plus mass of previously formed stars), MGF $= M_{\rm H_2} / (M_* + M_{\rm H_2})$, is also an important parameter that can be used to estimate e-folding times for the growth of galaxy stellar mass and depletion times for exhaustion of the fuel for star formation.  In the past, uncertainties in computing $M_*$ were often larger than the uncertainties in computing $M_{\rm H_2}$, but both estimates have since improved due largely to the extensive multi-wavelength data now routinely available for the GOALS LIRGs, as well as improvements in stellar evolution codes \citep[e.g.][]{Leitherer:1999aa}. 
This allows us to properly study how the MGF and infrared luminosity vary with merger stage for the first time.

\begin{table*}[ht]
\caption[]{Galaxy Properties for the GOALS Northern LIRG Sample } 
\centering
	\label{TAB:Properties}
\begin{tabular}{l l c c c r c c c c c c c c}

\hline
\hline

	  Name & Name & \multicolumn{3}{|c|}{RA (2000)} &\multicolumn{3}{|c|}{DEC (2000)}  & Tel. &  z & log($L_{\rm{IR}}$) &  log($M_*$) &\multicolumn{2}{|c|}{log($M_{\rm{H}_{2}}$)} \\
	  IRAS FSC  &  Common & hh &  mm &  ss.s &  $^{\circ}$ & $^{\prime}$ & $^{\prime\prime}$ &    &      &   ($L_\odot$)                 &  ($M_\odot$)   & ($M_\odot$) & Ref  \\

\hline
F00085-1223	&	NGC0034	&	00	&	11	&	06.6	&	-12	&	06	&	26	&	HST	&	 	0.020	&	11.48&	10.58	&	-	&	\\
F00163-1039	&	MCG-02-01-051	&	00	&	18	&	50.5	&	-10	&	22	&	09	&	HST	&	 	0.027	&	11.46&	10.65	&	-	&	\\
F00402-2349	&	NGC0232	&	00	&	42	&	45.8	&	-23	&	33	&	41	&	PS1	&	 	0.023	&	11.43&	10.84	&	10.18	&	5\\
F01053-1746	&	IC1623	&	01	&	07	&	47.2	&	-17	&	30	&	25	&	HST	&	 	0.020	&	11.66&	10.65	&	10.51	&	6\\
F01076-1707	&	MCG-03-04-014	&	01	&	10	&	09.0	&	-16	&	51	&	10	&	HST	&	 	0.033	&	11.63&	10.85	&	10.34	&	5\\
F01173+1405	&	CGCG436-030	&	01	&	20	&	02.7	&	14	&	21	&	43	&	HST	&	 	0.031	&	11.68&	10.56	&	-	&	\\
F01364-1042	&	IRASF01364	&	01	&	38	&	52.9	&	-10	&	27	&	11	&	HST	&	 	0.048	&	11.79&	10.45	&	10.18	&	5\\
F01417+1651	&	IIIZw035	&	01	&	44	&	30.4	&	17	&	06	&	05	&	HST	&	 	0.027	&	11.62&	10.25	&	9.93	&	6\\
F01484+2220	&	NGC0695	&	01	&	51	&	14.2	&	22	&	34	&	57	&	HST	&	 	0.032	&	11.69&	11.03	&	10.44	&	2 \\
F02281-0309	&	NGC0958	&	02	&	30	&	42.8	&	-02	&	56	&	20	&	PS1	&	 	0.019	&	11.22&	11.14	&	10.22	&	6\\
F02401-0013	&	NGC1068	&	02	&	42	&	40.7	&	00	&	00	&	48	&	PS1	&	 	0.004	&	11.40&	10.56	&	9.86	&	2\\
F02435+1253	&	UGC02238	&	02	&	46	&	17.5	&	13	&	05	&	44	&	PS1	&	 	0.022	&	11.39&	10.55	&	10.21	&	6 \\
F02512+1446	&	UGC02369	&	02	&	54	&	01.8	&	14	&	58	&	25	&	HST	&	 	0.031		&	11.60&	11.03	&	-	&	\\
F03359+1523	&	IRASF03359	&	03	&	38	&	46.7	&	15	&	32	&	55	&	HST	&	 	0.035	&	11.51&	10.29	&	10.43	&	6 \\
F04097+0525	&	UGC02982	&	04	&	12	&	22.5	&	05	&	32	&	51	&	UH 	&	 	0.018	&	11.20&	10.54	&	-	&	\\
F04191-1855	&	ESO550-IG025	&	04	&	21	&	20.0	&	-18	&	48	&	48	&	HST	&	 	0.032	&	11.50&	10.93	&	-	&	\\
F04315-0840	&	NGC1614	&	04	&	33	&	59.9	&	-08	&	34	&	44	&	HST	&	 	0.016	&	11.61&	10.63	&	10.09	&	6\\
F05189-2524	&	IRASF05189	&	05	&	21	&	01.5	&	-25	&	21	&	45	&	HST	&	 	0.043	&	12.13&	10.91	&	10.43	&	6 \\
F08354+2555	&	NGC2623	&	08	&	38	&	24.1	&	25	&	45	&	17	&	HST	&	 	0.019	&	11.58&	10.47	&	9.83	&	1 \\
F08572+3915	&	IRASF08572	&	09	&	00	&	25.4	&	39	&	03	&	54	&	HST	&	 	0.058	&	12.17&	10.29	&	9.84	&	4\\
F09126+4432	&	UGC04881	&	09	&	15	&	55.1	&	44	&	19	&	54	&	HST	&	 	0.039	&	11.70&	10.97	&	10.52	&	6 \\
F09320+6134	&	UGC05101	&	09	&	35	&	51.7	&	61	&	21	&	11	&	HST	&	 	0.039	&	12.00&	10.93	&	10.56	&	6\\
F09333+4841	&	MCG+08-18-013	&	09	&	36	&	37.2	&	48	&	28	&	28	&	UH 	&	 0.026	&	11.33&	9.56	&	-	&	\\
F09437+0317	&	IC0563-4	&	09	&	46	&	20.7	&	03	&	03	&	30	&	UH 	&	 	0.020	&	11.28&	10.87	&	10.32	&	6 \\
F10015-0614	&	NGC3110	&	10	&	04	&	02.1	&	-06	&	28	&	29	&	UH 	&	 	0.017	&	11.41&	10.83	&	10.21	&	6 \\
F10173+0828	&	IRASF10173	&	10	&	20	&	00.2	&	08	&	13	&	34	&	HST	&	 	0.049	&	11.79&	10.33	&	10.41	&	6 \\
F10565+2448	&	IRASF10565	&	10	&	59	&	18.1	&	24	&	32	&	34	&	HST	&	 	0.043	&	12.05&	10.87	&	10.4	&	6\\
F11011+4107	&	MCG+07-23-019	&	11	&	03	&	53.2	&	40	&	50	&	57	&	HST	&	 	0.035	&	11.63&	10.63	&	10.24	&	6\\
F11186-0242	&	CGCG011-076	&	11	&	21	&	12.3	&	-02	&	59	&	03	&	UH 	&	 	0.025	&	11.38&	10.75	&	10.08	&	6 \\
F11231+1456	&	IC2810	&	11	&	25	&	47.3	&	14	&	40	&	21	&	HST	&	 	0.034	&	11.64&	10.99	&	10.32	&	6\\
F11257+5850	&	NGC3690	&	11	&	28	&	32.3	&	58	&	33	&	44	&	HST	&	 	0.010	&	11.89&	10.76	&	10.42	&	6 \\
F12112+0305	&	IRASF12112	&	12	&	13	&	46.0	&	02	&	48	&	38	&	HST	&	 	0.073		&	12.33&	10.76	&	10.68	&	6\\
F12224-0624	&	IRASF12224	&	12	&	25	&	03.9	&	-06	&	40	&	53	&	UH 	&	 	0.026		&	11.32&	9.95	&	-	&	\\
F12540+5708	&	UGC08058	&	12	&	56	&	14.2	&	56	&	52	&	25	&	HST	&	 	0.042		&	12.53&	11.57	&	10.55	&	6\\
F12590+2934	&	NGC4922	&	13	&	01	&	24.9	&	29	&	18	&	40	&	UH 	&	 	0.024		&	11.33&	10.97	&	-	&	\\
F13001-2339	&	ESO507-G070	&	13	&	02	&	52.3	&	-23	&	55	&	18	&	HST	&	 	0.022	&	11.53&	10.78	&	10.03	&	5 \\
F13126+2453	&	IC0860	&	13	&	15	&	03.5	&	24	&	37	&	08	&	UH 	&	 	0.011		&	11.10&	10.12	&	9.08	&	6 \\
F13136+6223	&	VV250a	&	13	&	15	&	35.1	&	62	&	07	&	29	&	HST	&	 	0.031		&	11.77&	10.39	&	10.19	&	6\\
F13182+3424	&	UGC08387	&	13	&	20	&	35.3	&	34	&	08	&	22	&	HST	&	 	0.023		&	11.72&	10.59	&	9.93	&	6 \\
F13188+0036	&	NGC5104	&	13	&	21	&	23.1	&	00	&	20	&	33	&	UH   	&	 	0.019		&	11.25&	10.81	&	9.95	&	6 \\
F13197-1627	&	MCG-03-34-064	&	13	&	22	&	24.5	&	-16	&	43	&	43	&	UH 	&	 	0.017		&	11.19&	10.75	&	-	&	\\
F13229-2934	&	NGC5135	&	13	&	25	&	44.1	&	-29	&	50	&	01	&	UH 	&	 	0.014		&	11.29&	10.97	&	10.12	&	6\\
F13362+4831	&	NGC5256	&	13	&	38	&	17.5	&	48	&	16	&	37	&	HST	&	 	0.028		&	11.52&	11.01	&	10.21	&	2\\
F13373+0105	&	NGC5257-8	&	13	&	39	&	55.0	&	00	&	50	&	07	&	HST	&	 	0.023		&	11.63&	11.23	&	-	&	\\
F13428+5608	&	UGC08696	&	13	&	44	&	42.1	&	55	&	53	&	13	&	HST	&	 	0.038		&	12.18&	10.96	&	10.33	&	6\\
F14179+4927	&	CGCG247-020	&	14	&	19	&	43.3	&	49	&	14	&	12	&	UH 	&	 	0.026		&	11.35&	10.45	&	-	&	\\
F14348-1447	&	IRASF14348	&	14	&	37	&	38.4	&	-15	&	00	&	23	&	HST	&	 	0.083		&	12.37&	11.02	&	10.84	&	6\\
F14547+2449	&	VV340a	&	14	&	57	&	00.7	&	24	&	37	&	03	&	HST	&	 	0.034		&	11.79&	10.83	&	-	&	\\
F15107+0724	&	CGCG049-057	&	15	&	13	&	13.1	&	07	&	13	&	32	&	UH 	&	 	0.013		&	11.33&	10.02	&	9.53	&	6\\
F15163+4255	&	VV705	&	15	&	18	&	06.3	&	42	&	44	&	41	&	HST	&	 	0.040		&	11.88&	10.86	&	10.33	&	6\\
F15250+3608	&	IRASF15250	&	15	&	26	&	59.4	&	35	&	58	&	38	&	HST	&	 	0.055		&	12.07&	10.61	&	-	&	\\
F15327+2340	&	UGC09913	&	15	&	34	&	57.1	&	23	&	30	&	11	&	HST	&	 	0.018		&	12.24&	10.81	&	10.34	&	6 \\
F16104+5235	&	NGC6090	&	16	&	11	&	40.7	&	52	&	27	&	24	&	HST	&	 	0.029		&	11.55&	10.73	&	10.21	&	2\\
F16284+0411	&	CGCG052-037	&	16	&	30	&	56.5	&	04	&	04	&	58	&	UH 	&	 	0.024		&	11.45&	10.72	&	-	&	\\
F16577+5900	&	NGC6286	&	16	&	58	&	31.4	&	58	&	56	&	10	&	UH 	&	 	0.018		&	11.42&	10.76	&	10.03	&	2\\
F17132+5313	&	IRASF17132	&	17	&	14	&	20.0	&	53	&	10	&	30	&	HST	&	 	0.051		&	11.92&	10.89	&	10.61	&	2 \\
F22287-1917	&	ESO602-G025	&	22	&	31	&	25.5	&	-19	&	02	&	04	&	UH 	&	 	0.025		&	11.34&	10.82	&	-	&	\\
F22491-1808	&	IRASF22491	&	22	&	51	&	49.3	&	-17	&	52	&	23	&	HST	&	 	0.078		&	12.19&	10.71	&	10.49	&	6\\
F23007+0836	&	NGC7469	&	23	&	03	&	15.6	&	08	&	52	&	26	&	HST	&	 	0.016		&	11.58&	11.05	&	10.31	&	1\\
F23024+1916	&	CGCG453-062	&	23	&	04	&	56.5	&	19	&	33	&	08	&	UH 	&	 	0.025		&	11.37&	10.62	&	-	&	\\
F23135+2517	&	IC5298	&	23	&	16	&	00.7	&	25	&	33	&	24	&	HST	&	 	0.027		&	11.53&	10.76	&	9.95	&	6 \\
F23157-0441	&	NGC7592	&	23	&	18	&	22.2	&	-04	&	24	&	58	&	UH 	&	 	0.024		&	11.39&	10.73	&	10.34	&	6\\
F23254+0830	&	NGC7674	&	23	&	27	&	56.7	&	08	&	46	&	45	&	HST	&	 	0.029		&	11.51&	11.17	&	10.12	&	1\\
F23488+1949	&	NGC7771	&	23	&	51	&	24.9	&	20	&	06	&	43	&	UH 	&	 	0.014		&	11.35&	11.08	&	10.01	&	2\\
F23488+2018	&	MRK0331	&	23	&	51	&	26.8	&	20	&	35	&	10	&	HST	&	 	0.018		&	11.50&	9.67	&	-	&	\\
\hline
\vspace{0.05truein}
\end{tabular}

Galaxy properties of the sample: Column (1) $IRAS$ Faint Source Catalog name; Column (2) common name; Column (3) right ascension; Column (4) declination; Column (5) Telescope used to obtain galaxy $I$ band image: $HST$---Hubble Space Telescope, PS1---Pan-STARRS1 Survey, UH---UH 2.2m; Column (6) Redshift; Column (7)   $L (8 - 1000\mu m)$  \citep{U:2012aa}; Column (8)  $H$ band galaxy mass using Saltpeter IMF  \citep{U:2012aa}; Column (9) molecular gas mass; Column (10) Reference for molecular gas mass:\  1---\citet{Sanders:1985aa}, 2---\citet{Sanders:1986aa}, 3---\citet{Mirabel:1988aa}, 4---\citet{Sanders:1989aa}, 5---\citet{Mirabel:1990aa}, 6---\citet{Sanders:1991aa}
\end{table*}

The paper is organized as follows:  Section \ref{SEC:Sample} outlines the sample selection. Section  \ref{SEC:Data} summarizes the data used in this paper.  Section \ref{SEC:Analysis} presents our analysis of these data including the introduction of a new visual classification scheme  used to determine galaxy morphology, presentation of galaxy stellar masses and the calculation of molecular gas masses using previously published large-measurements of total CO(1-0) emission. The results of our visual classification for all 65 objects as well as the derived MGFs are given in Section \ref{SEC:Results}. Comparisons of morphology and MGFs to infrared luminosities and merger stage are in Section \ref{SEC:Discussion}. Our conclusions are summarized in Section \ref{SEC:Conclusions}. Throughout this paper, we adopt a flat model of the universe with a Hubble constant $H_0 = 70$ km s$^{-1}$ Mpc$^{-1}$, and $\Omega_{\rm m}$ = 0.28, and $\Omega_{\Lambda}$ = 0.72 \citep{Komatsu:2009aa}.

\section{The Northern LIRG Sample}
\label{SEC:Sample}

The ``Northern" sample of 65 LIRGs discussed in the paper was originally defined by \citet{Ishida:2004aa} to select those LIRGs in the RBGS that were visible from Mauna Kea (i.e. $\delta > -30^{\circ}$) and with $|b| > 30^{\circ}$) to minimize galactic extinction. The RBGS is an all-sky,  complete flux-limited survey of extra-galactic objects with total 60 $\mu$m flux greater than 5.24 Jy.  Of the 629 objects in the all-sky RBGS, 203 are LIRGs, i.e. log$(L_{\rm IR} /L_\odot) > 11.0$, and 90 of these LIRGs are in our Northern region of the sky.  

\citet{Ishida:2004aa} used the UH 2.2m telescope to observe a total of 65 objects from the complete flux-limited sample of 90 objects in the Northern LIRG Sample. The northern observations are complete above log$(L_{\rm IR} /L_\odot) > 11.54$ with all 38 objects being observed. For the 42 lower luminosity LIRGs with log$(L_{\rm IR} /L_\odot) = 11.01 - 11.54$, approximately half ($27/52 = 52$\%) were observed by \citet{Ishida:2004aa}. The availability of deep $I$-band wide-field imaging \citep{Ishida:2004aa} and total galaxy stellar masses \cite{U:2012aa} for the Northern LIRG Sample, was critical for determining the morphology classifications and computing the molecular gas fractions, respectively, that are presented in the current paper.  Table \ref{TAB:Properties} gives the object names, data references, and lists the general galaxy properties for all 65 objects in the Northern LIRG sample.     

This paper is the third in a series of papers describing the multi-wavelength properties of the Northern LIRG  Sample.  The original optical imaging data were presented by \citet{Ishida:2004aa}.  A second paper \citep{U:2012aa} presented complete spectral energy distributions (SEDs) using the large amount of multi-wavelength data now available in GOALS, and used these data to compute more accurate galaxy stellar masses ($M*$) and infrared luminosities ($L_{\rm IR}$).  The current paper presents a more detailed analysis of galaxy morphology along with an analysis of the variation of galaxy properties. A companion paper (fourth in this series) compares the visual classifications with classifications derived using automated methods (e.g. $Gini$, $M20$, $C$ompactness, $A$ssymetry, $S$moothness)(K. Larson et al., in\ preparation).  The fifth and final paper in this series (K. Larson et al., in preparation) will present a detailed analysis of the radial distribution of optical colors and mean stellar ages for individual galaxies as well as mean colors as a function of merger stage. 

\section{Data}
\label{SEC:Data}
We use optical imaging data from the GOALS $HST$ sample \citep[{PID: 10592, PI: Evans; }[ ]{Kim:2013aa} 
Maunakea \citep{Ishida:2004aa}, and Pan-STARRS1 observations.
Twenty-two LIRGS in our sample have optical $I$-band images obtained with the UH 2.2m telescope on Maunakea by \citet{Ishida:2004aa}. The images taken with the UH 2.2m telescope have a total $900$\,s exposure time with a plate scale of  $0.2''$pixel$^{-1}$ and were observed in seeing of $0.5-1.2''$. The seeing corresponds to a physical scale of $0.2-0.6$\,kpc at the distance of the galaxies.
Forty-three (U)LIRGs have $HST$ data taken with the $F814W$ filter using the ACS camera \citep{Kim:2013aa}. The $HST$ data have $720$\,s exposures at a plate scale of $0.05''$ with seeing of $0.1 - 0.15''$ corresponding to an average physical scale of 0.1\,kpc.
Pan-STARRS1 Survey \citep[PS1:][]{Schlafly:2012aa,Tonry:2012aa,Magnier:2013aa} data were used for four galaxies whose original UH 2.2m data were insufficient due to either high background noise or a limited field of view. The PS1 mosaic images have total exposure times ranging from 990 to 1260\,s with a  drizzled plate scale of $0.25''$ and a seeing of $1''$ which corresponds to a physical scale of $0.08-0.5$\,kpc.

Table \ref{TAB:Properties} also lists the total mass of molecular gas for each galaxy using previously published values of $M_{\textrm{H}_{2}}$ computed from large aperture millimeterwave observations of the CO(1-0) emission line, where the previous values have been updated to reflect our assumed cosmology and our adopted value for the CO$\rightarrow$$M_{\textrm{H}_{2}}$ conversion factor, $X_{\rm CO} = 3.0 \times 10^{20}$\,H$_2$\,cm$^{-2}$(K\,km\,s$^{-1})^{-1}$.


\section{Analysis}
\label{SEC:Analysis}


\subsection{Visual Morphological Classification Scheme}
\label{SEC:VisClass}

The GOALS team has previously classified the morphologies of galaxies
in the Northern LIRG Sample \citep[e.g][]{Surace:1998aa, Haan:2011aa, Kim:2013aa, Stierwalt:2013aa} 
using a variety of merger classifications.
We chose to improve upon these previous classifications for several
reasons.
Firstly, our sample extends to lower luminosities (log$(L_{\rm IR} /L_\odot) = 11.1 - 11.63$) than the sources analyzed by 
\citet{Surace:1998aa}, \citet{Haan:2011aa}, and \citet{Kim:2013aa}, so 37\% of our sample was not classified in
their work.
These lower luminosities include a large number of single systems  
which were relatively rare in previous samples.
Second, we use high resolution $HST$ and UH\,2.2m $I$-band images to
classify galaxies, while \citet{Stierwalt:2013aa} used $Spitzer$/IRAC $3.6\mu$m
images.
The IRAC images have $2^{\prime\prime}$ resolution, making it difficult to
distinguish close double nuclei.
Furthermore, the relatively shallow ($t_{\rm exposure} = 120$\,s) IRAC images precluded the
identification of interacting minor companions or fainter tidal
disturbances.
Twenty three of the objects in our sample use ground-based optical imaging to improve upon \citet{Stierwalt:2013aa} classifications which have on average 3 to 4 times better resolution than the IRAC data. 

Finally, the classification scheme used by \citet{Kim:2013aa}and
\citet{Surace:1998aa} placed considerable emphasis on higher
infrared luminosities characteristic of end-stage mergers.
Examination of galaxy encounters using Identikit \citep{Barnes:2009aa}
has allowed us to explore the parameter space of galaxy collisions and
mergers.
This has led us to a different perspective on classification and we
have simplified the classification of the most important merger
stages.

We visually classify our galaxies using a new scheme that accommodates
a mixture of minor mergers, major mergers, and galaxies which 
show no sign of current interaction, and allows for ambiguities due
to projection effects.
When two separated galaxies are present we used velocities from the NASA/IPAC Extragalactic Database (NED) to distinguish between true close pairs and projection effects. 
On large scales, galaxy pairs are required to have line of sight velocities 
differing by less than $250$\,km/s and projected separations of less than 
$75$\,kpc to be considered as interacting.
If needed velocity information is missing causing the classification to be unclear, we define the merger class as ambiguous (amb).
On small scales, angular resolution and optical depth effects limit our ability to distinguish
between single and double nuclei; therefore, taking the redshift range (0.004 $\le$ z $\le$ 0.083) of our
sample into account, only systems with projected nuclear separations
of more than $2$\,kpc are considered to have multiple nuclei. 
The minimum nuclear separation of $2$\,kpc is conservative seeing limit chosen to be greater than the maximum nuclear full width half maximum (FWHM) of the data.
This is larger than the average seeing of the data since the nuclear regions are often slightly extended and not perfect point sources. Therefore, even extended nuclei should be visibly discernible.

\begin{figure}[!]
	\centering
	\includegraphics[width=0.42\textwidth]{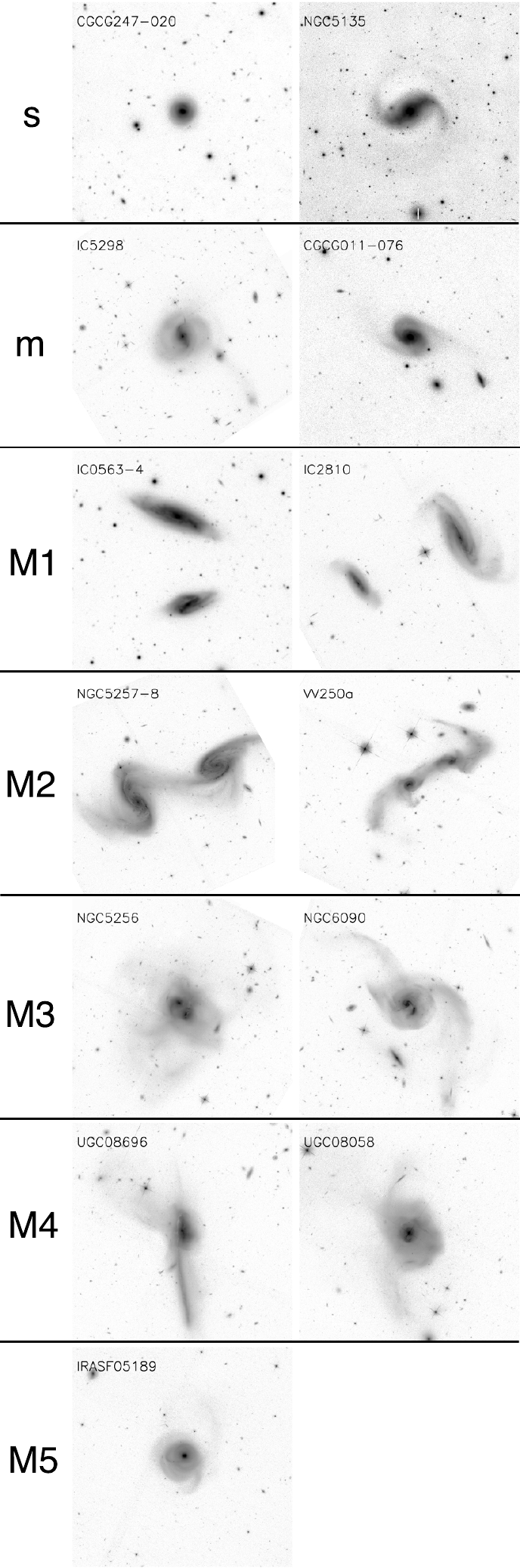}
	\caption{Visual morphological classification scheme}
	\label{FIG:Morph_fig}
\end{figure}
\begin{figure*}[!]
	\centering
	\includegraphics[width=0.95\textwidth]{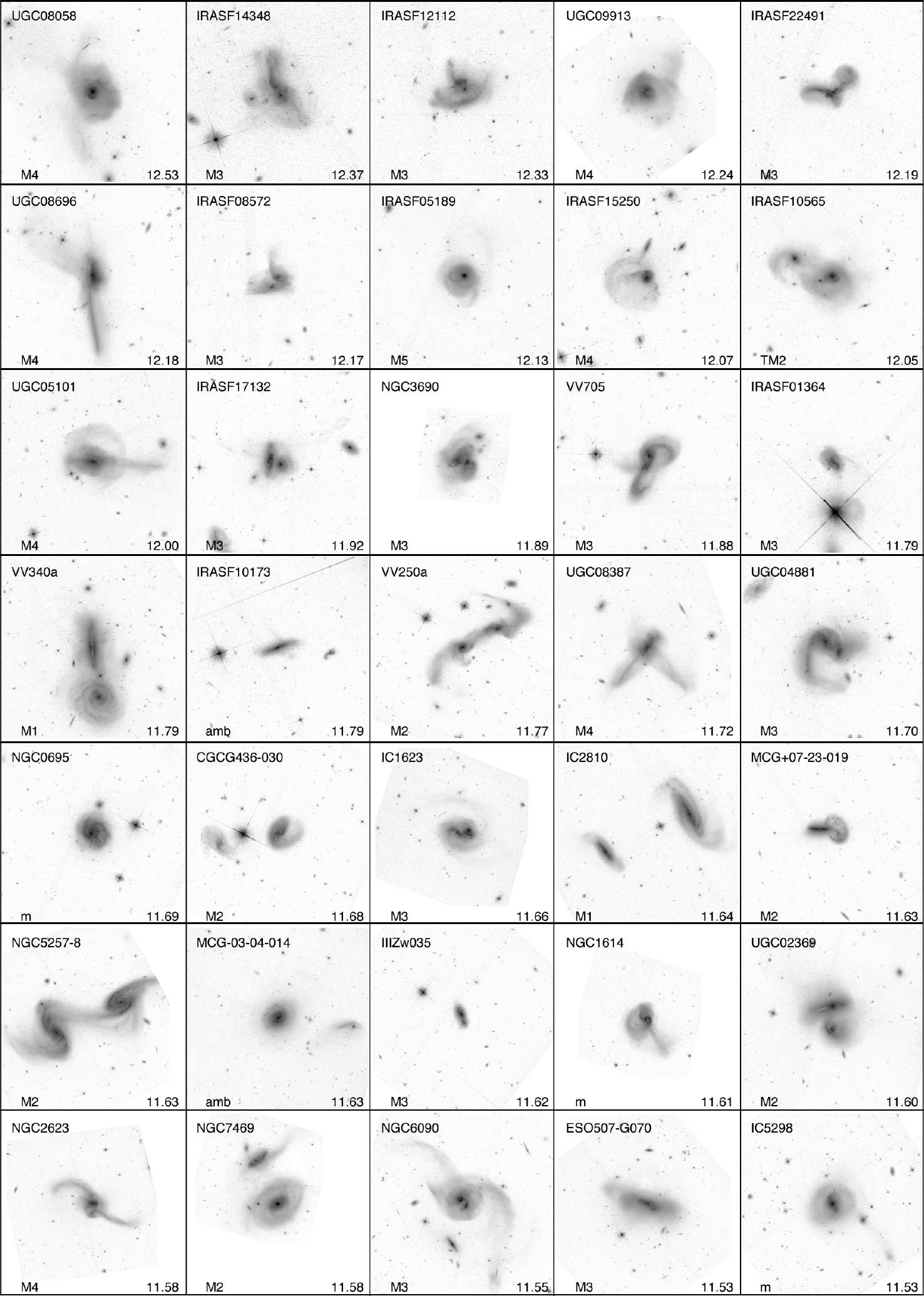}
	\caption{Continued on next page.}
\end{figure*}
\setcounter{figure}{1}
\begin{figure*}[!]
	\centering
	\includegraphics[width=0.95\textwidth]{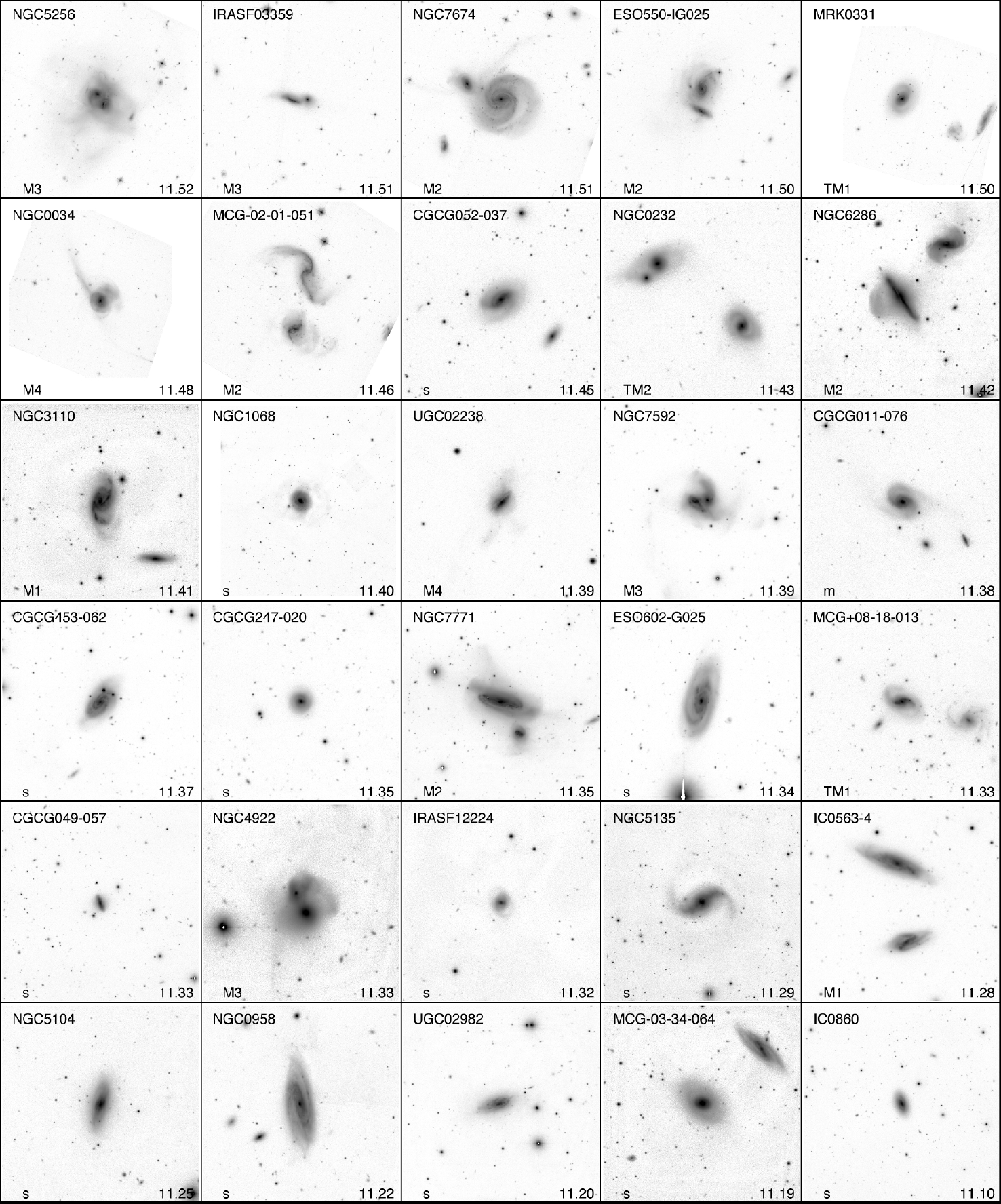}
	\caption{$I$-band images ($FOV = $ 100\,kpc$\times$100\,kpc) of all 65 galaxies in the Northern LIRG sample, in order of decreasing infrared luminosity.  Images were obtained from $HST$, UH 2.2m, and PS1. Galaxy merger stage and infrared luminosity are given in the bottom left and right corners, respectively, of each panel. All images are oriented with North as up. }
	\label{FIG:Ish_images2}
\end{figure*}
Guided by these parameters, we classify each object as follows:

\begin{enumerate}
\begin{item}[\textbf{s}] Single galaxy:\  No current sign of an interaction
or merger event.
\end{item}
\begin{item}[\textbf{m}] Minor merger:\  Interacting pairs with estimated
mass ratios $> 4:1$. 
\end{item}
\begin{item}[\textbf{M1}] Major merger - stage 1:\  Galaxy pairs with $\Delta V < 250$ km/s and $n_{\rm sep} < 75$ kpc, which
have no prominent tidal features. These
galaxies appear to be on their initial approach.
\end{item}
\begin{item}[\textbf{M2}] Major merger - stage 2:\  Interacting galaxy
pairs with obvious tidal bridges and tails \citep{Toomre:1972aa}
 or other disturbances consistent with having already undergone
a first close passage.
\end{item}
\begin{item}[\textbf{M3}] Major merger - stage 3:\  Merging galaxies
with multiple nuclei. These systems have distinct nuclei in disturbed,
overlapping disks, along with visible tidal tails.
\end{item}
\begin{item}[\textbf{M4}] Major merger - stage 4: Galaxies with apparent single
nuclei and obvious tidal tails. The galaxy nuclei have $n_{\rm sep} \lesssim 2$\,kpc.
\end{item}
\begin{item}[\textbf{M5}] Major merger - stage 5:\  Galaxies which appear to be evolved merger
remnants. These galaxies have diffuse envelopes which may exhibit
shells or other fine structures \citep{Schweizer:1992aa} and a single, possibly
off-center nucleus. These merger remnants no longer have bright tidal tails.
\end{item}
\end{enumerate}

A few ($\sim 5$\%) of our objects appear to contain three distinct nuclei.
In parallel with the scheme outlined above, they are classified as TM1,
TM2, or TM3 depending on the earliest major merger interaction stage
involved.  For example, a hierarchical triplet consisting of a close M3
pair interacting with a separate galaxy displaying a prominent tidal
tail would be classified as TM2. 
Triple systems are explicitly identified when included in figures 
since it is unclear how the third nuclei affects the galaxy properties.

Previous studies by \citet{Surace:1998aa}, \citet{Haan:2011aa}, and \citet{Kim:2013aa} used
a classification scheme containing six major merger stages.
This scheme put greater emphasis on the brightness of the tidal
tails, and included an intermediate stage between our M2 and M3
classifications, described as `galaxies in a common envelope.'
However, tail brightness is projection-dependent: tails are thin,
curving ribbons of tidal material \citep{Toomre:1972aa}, which
typically appear much brighter when viewed edge-on.
Projection effects also complicate the interpretation of some `common
envelope' objects, which may be either be M2 interacting pairs seen in
projection, or M3 mergers with double nuclei.
Our visual classifications can still be affected by projection
effects; for example, a pair of well-separated interacting galaxies
(merger stage M2) will appear to be in merger stage M3 if viewed along
an unfavorable line of sight.
However, we believe that our simplified system will prove robust to
projection effects, while retaining enough morphological
discrimination to identify physically distinct merger stages.

Examples of the different visual galaxy classifications are shown in figure \ref{FIG:Morph_fig}.


\subsection{Molecular Gas and Stellar Masses}
 
There have been several large observing programs to study the total molecular gas content of LIRGs, and the GOALS objects in particular, using single-dish telescopes with beam-size larger than the optical diameter of the observed host galaxy \citep[e.g.][see also Table \ref{TAB:Properties} for additional references]{Sanders:1991aa}.   Forty-seven of the 65 targets in our Northern LIRG sample have CO observations from these previous programs. 
The published values of $M_{\rm H_2}$ have been adjusted to account for our adopted cosmology and adopted value for the CO$\rightarrow$$M_{\textrm{H}_{2}}$ conversion factor, $X_{\rm CO}  = 3.0 \times 10^{20}$ H$_2$ cm$^{-2}$(K km s$^{-1}$)$^{-1}$, and are listed in Table \ref{TAB:Properties} along with the original reference.   
The given stellar masses, $M_*$, were calculated using a Saltpeter IMF by \citet{U:2012aa}, and are listed for all 65 of our sources in Table \ref{TAB:Properties}.   

\section{Results}
\label{SEC:Results}

\subsection{Visual Classifications}

Most of the 65 galaxies in our sample have bright nuclei, tidal bridges, and tidal tails allowing them to be visually classified into merger stages.  Only 2 sources were ambiguous, 4 are triple nuclei major merging systems, 4 are minor mergers, and 14 are single galaxies. Our sample has a total of 45 major merging systems with 4 triple systems, 3 M1, 11 M2, 17 M3, 9 M4, and 1 M5 galaxy. Figure \ref{FIG:Ish_images2} shows the $I$-band images for all galaxies arranged by decreasing infrared luminosity and labeled by interaction class. Further details on the individual classifications for all galaxies are given in the Appendix.

\subsection{Visual Classifications versus Infrared Luminosity.}

\begin{figure}[t!]
	\includegraphics[width= 0.49\textwidth]{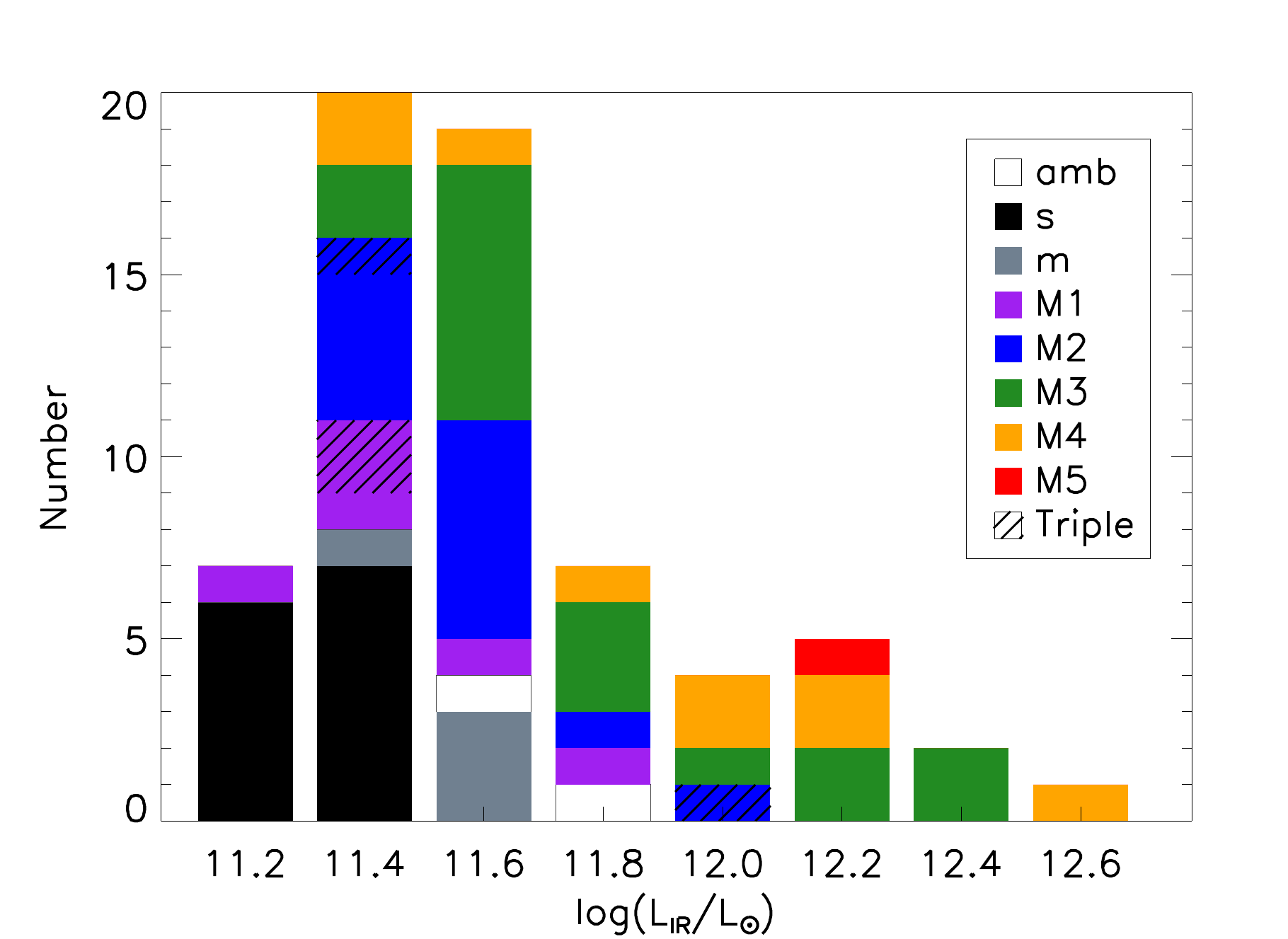}
	\caption{Distribution of morphology types with infrared luminosity. Normal galaxies (black), minor merger (gray), major separated pair (purple), major interacting pair (blue), major merging double nucleus (green), major merging single nucleus (orange), diffuse merger remnant (red), ambiguous (white), triple system (hashed).  Each bin is labeled by its central infrared luminosity and is 0.2 log(L$_{\tiny{\textrm{IR}}}$/L$_\odot$) wide. }
	\label{FIG:mstage_hist}
\vspace{0.1truein}
\end{figure}

We investigate the dependence of infrared luminosity with merger stage. We divide the infrared luminosity into bins of  0.2 dex in log $\ L_{\rm IR}$ giving an average of 8 galaxy systems per bin. Our bin size is limited by the scarcity of sources at the highest infrared luminosities making it unreasonable to decrease the bin size any further.
Figure\,$ $\,\ref{FIG:mstage_hist} shows the varying contributions at each luminosity bin from each merger stage. 

All galaxies above an infrared luminosity $10^{11.5}\ L_\odot $ are interacting systems. Furthermore, all galaxies above an infrared luminosity of $10^{11.7}\ L_\odot $  are major mergers and almost all ultra luminous infrared galaxies (ULIRGs: $> 10^{12.0}\ L_\odot $) are late-stage mergers (M3, M4, M5).
The only ULIRG that is not a late-stage merger is IRAS F10565 which is a triple nuclei merging system and classified as a TM2. 
This shows that it is possible for galaxies to achieve infrared luminosity of up to $10^{11.5}\ L_\odot$ without any visible sign of a galaxy interaction, but a major merging event is required to boost the infrared luminosity to the ULIRG stage.

\subsection{Molecular Gas Fraction versus Infrared Luminosity}

\begin{figure}[h]
	\includegraphics[width= 0.49\textwidth]{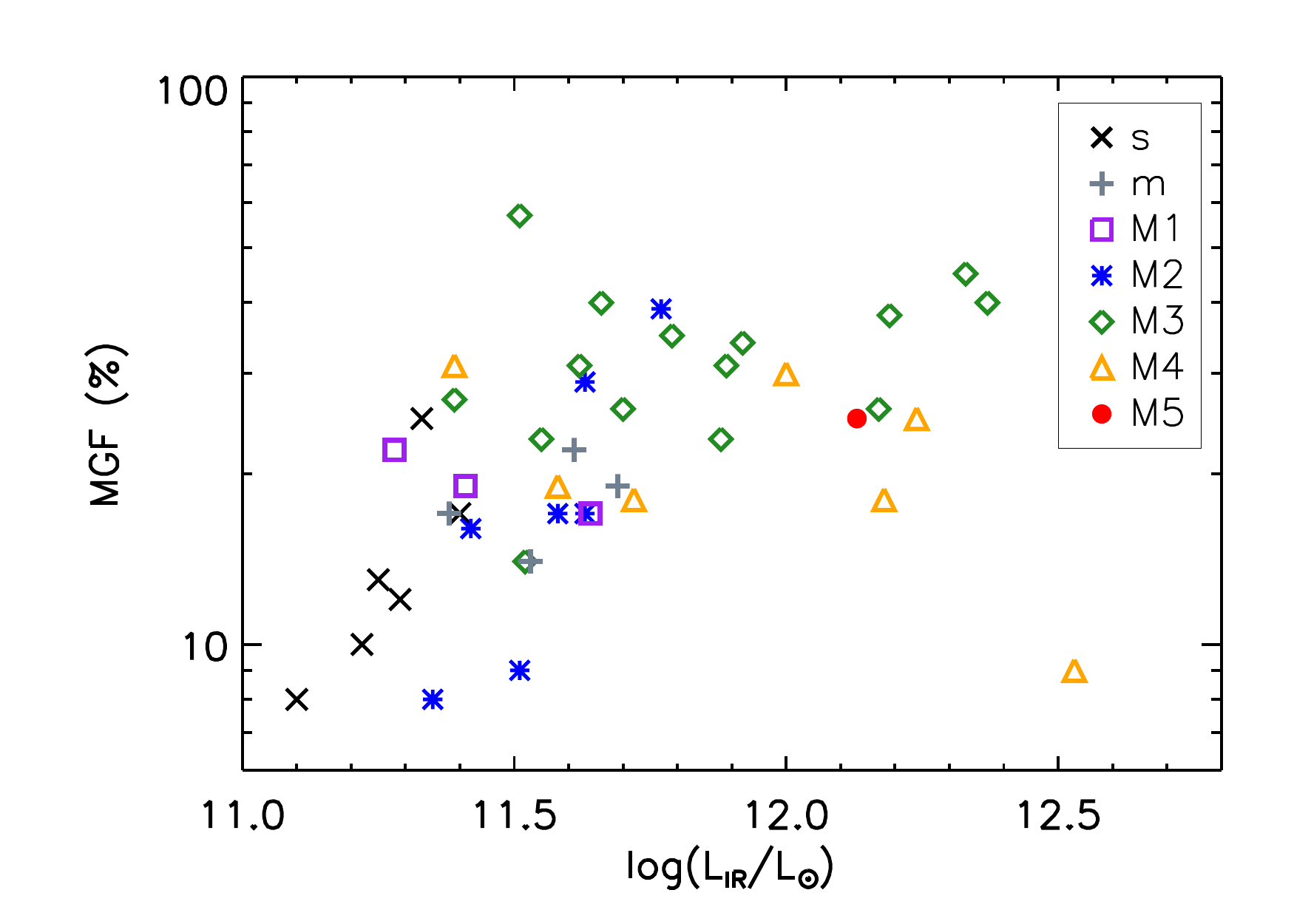}
	\caption{The molecular gas fraction (MGF),  $M_{\rm H_2} / (M_* + M_{\rm H_2})$, is given as a percentage versus infrared luminosity where each object is color coded according to assigned merger stage.}
	\label{FIG:LirGas}
\end{figure}

For the 47 galaxies with CO observations, we have computed the MGF,  $M_{\rm H_2} / (M_* + M_{\rm H_2})$, which are listed in Table \ref{TAB:Morph_params}.  Figure \ref{FIG:LirGas} shows a plot of the MGF versus infrared luminosity, where each object is also color coded to represent its visual classification stage. A fairly abrupt increase and flattening is observed in the MGF for nearly all objects above $L_{\rm IR} \sim 10^{11.5} L_\odot$. Compared to the mean MGF of 12.7$\%$ associated with LIRGs at lower infrared luminosity, LIRGs above $L_{\rm IR} \sim 10^{11.5} L_\odot$ have a mean MGF of 23.8$\%$. It is notable that this ``infrared luminosity threshold" is also where, above which, a majority of LIRGs are associated with major mergers. Not every galaxy in our sample has CO observations resulting in a completeness 75\% for the sample with a 88\% completeness for late-stage mergers (M3, M4, M5) and 61\% for early-stage mergers (M1, M2).

\begin{table*}[h!]
\caption[]{Galaxy Morphology and Molecular Gas Fraction in $L_{\rm IR}$ Order}
\label{TAB:Morph_params}
\centering
\begin{tabular}{l c c c c l}

\hline
\hline \\
Name  	& log($L_{\rm{IR}}$)	&  $n_{\rm sep}$	& Visual		&  MGF  &  Notes\footnotemark[1] \\
Common        &	($L_\odot$) 	&  (kpc)    			& Class	&     (\%)  &   \\
\hline \\
UGC08058 		&	12.53	&	 0.64 	&	 M4 	&	 9  	&		\\
IRASF14348 		&	12.37	&	 5.47 	&	 M3 	&	 40 	&		\\
IRASF12112 		&	12.33	&	 4.28 	&	 M3 	&	 45  	&		\\
UGC09913 		&	12.24	&	 0.72 	&	 M4 	&	 25  	&		\\
IRASF22491 		&	12.19	&	 2.68 	&	 M3 	&	 38  	&		\\
UGC08696 		&	12.18	&	 0.77 	&	 M4 	&	 18  	&		\\
IRASF08572 		&	12.17	&	 6.62 	&	 M3 	&	 26  	&		\\
IRASF05189 		&	12.13	&	 0.19 	&	 M5 	&	 25  	&		\\
IRASF15250 		&	12.07	&	 1.27 	&	 M4 	&	 -  	&		\\
IRASF10565 		&	12.05	&	 24.75 	&	 TM2 &	 25 	&	 Triple system	\\
UGC05101 		&	12.00	&	 0.40 	&	 M4 	&	 30  	&		\\
IRASF17132 		&	11.92	&	 10.49 	&	 M3 	&	 34  	&		\\
NGC3690 		&	11.89	&	 9.59 	&	 M3 	&	 31  	&		\\
VV705 			&	11.88	&	 6.26 	&	 M3 	&	 23  	&		\\
IRASF10173 		&	11.79	&	 - 		&	 amb &	 55  	&	 M4 or M2 	\\
IRASF01364 		&	11.79	&	 2.11 	&	 M3 	&	 35  	&		\\
UGC04881 		&	11.79	&	 11.15 	&	 M3 	&	 26  	&		\\
VV250a 			&	11.77	&	 42.48 	&	 M2 	&	 39  	&		\\
UGC08387 		&	11.72	&	 1.52 	&	 M4 	&	 18  	&		\\
VV340a 			&	11.70	&	 27.67 	&	 M1 	&	 -  	&		\\
NGC0695 		&	11.69	&	 16.50 	&	 m  	&	 19  	&		\\
CGCG436-030 		&	11.68	&	 34.51 	&	 M2 	&	 -  	&		\\
IC1623 			&	11.66	&	 6.42 	&	 M3 	&	 40  	&		\\
IC2810 			&	11.64	&	 52.02 	&	 M1 	&	 17 	&		\\
NGC1614 		&	11.63	&	 2.96 	&	 m  	&	 22  	&		\\
NGC5257/8 		&	11.63	&	 39.49 	&	 M2 	&	 -  	&		\\
MCG+07-23-019 	&	11.63	&	 12.94 	&	 M2 	&	 29 	&		\\
IIIZw035 			&	11.62	&	 4.57 	&	 M3 	&	 31  	&		\\
NGC7469 		&	11.61	&	 26.24 	&	 M2 	&	17	&		\\
UGC02369 		&	11.60	&	 12.88 	&	 M2 	&	 -  	&		\\
MCG-03-04-014	&	11.58	&	 - 		&	 amb &	 21  	&	 M2 or m 	\\
NGC2623 		&	11.58	&	 0.20 	&	 M4 	&	 19  	&		\\
NGC6090 		&	11.55	&	 6.23 	&	 M3 	&	 23 	&		\\
IC5298 			&	11.53	&	 20.87 	&	 m  	&	 14  	&		\\
NGC5256 		&	11.53	&	 6.95 	&	 M3 	&	 14  	&		\\
NGC7674 		&	11.52	&	 19.38 	&	 M2 	&	 9  	&		\\
ESO507-G070 		&	11.51	&	 4.68 	&	 M3 	&	 15  	&		\\
IRASF03359 		&	11.51	&	 6.51 	&	 M3 	&	 57  	&		\\
ESO550-IG025 	&	11.50	&	 11.27 	&	 M2 	&	 - 	&		\\
NGC0034 		&	11.48	&	 0.15 	&	 M4 	&	 -  	&		\\
MCG-02-01-051 	&	11.46	&	 30.23 	&	 M2 	&	 - 	&		\\
CGCG052-037 		&	11.45	&	 1.13 	&	 s 	&	 -  	&		\\
NGC0232 		&	11.43	&	 56.53 	&	 TM2 &	 18 	&	 Triple system	\\
NGC3110 			&	11.42	&	 40.63 	&	 M1 	&	19	&		\\
NGC6286 		&	11.41	&	 51.99 	&	 M2 	&	 16 	&		\\
NGC1068 		&	11.40	&	 0.15 	&	 s 	&	 17  	&		\\
NGC7592 		&	11.39	&	 5.35 	&	 M3 	&	 27  	&		\\
UGC02238 		&	11.39	&	 1.14 	&	 M4 	&	 31 	&		\\
CGCG011-076 		&	11.38	&	 37.36 	&	 m  	&	 17 	&		\\
CGCG453-062 		&	11.37	&	 1.25 	&	 s 	&	 -  	&		\\
Mrk0331 			&	11.36	&	 41.02 	&	 TM1	&	 - 	&	 Triple system	\\
NGC7771 		&	11.35	&	 18.48 	&	 M2 	&	 8  	&		\\
CGCG247-020 		&	11.35	&	 0.52 	&	 s 	&	 -  	&		\\
MCG+08-18-013 	&	11.34	&	 52.23 	&	 TM1 &	 - 	&	 Triple system	\\
NGC4922 		&	11.33	&	 11.54 	&	 M3 	&	 - 	&		\\
CGCG049-057 		&	11.33	&	 1.02 	&	 s 	&	 25  	&		\\
ESO602-G025 		&	11.33	&	 0.79 	&	 s 	&	 -  	&		\\
IRASF12224 		&	11.32	&	 1.19 	&	 s 	&	 -  	&		\\
NGC5135 		&	11.29	&	 1.07 	&	 s 	&	 12  	&		\\
IC0563/4 			&	11.28	&	 40.35 	&	 M1 	&	 22 	&		\\
NGC5104 		&	11.25	&	 1.86 	&	 s 	&	 13  	&		\\
NGC0958 		&	11.22	&	 0.79 	&	 s 	&	 10  	&		\\
UGC02982 		&	11.20	&	 1.39 	&	 s 	&	 -  	&		\\
MCG-03-34-064 	&	11.19	&	 0.69 	&	 s 	&	 -  	&		\\
IC0860 			&	11.10	&	 0.62 	&	 s 	&	 8  	&		\\

\hline
\footnotetext[1]{Additional notes for the classification of each galaxy are given in the Appendix.}
\end{tabular}
\end{table*}

\section{Discussion}
\label{SEC:Discussion}
We have visually classified all 65 (U)LIRGs in our Northern LIRG sample. Our method of visual classification fully accounts for all possible interaction stages represented in our sample: s (single galaxies with no sign of interaction), m (minor merger, galaxies with a mass difference of $<$\,4:1), M1 (Major merger, separated galaxy pair), M2 (Major merger, interacting galaxy pair), M3 (Major merger, merging galaxy with two nuclei and tidal tails), M4 (Major merger, merging galaxy with a single nucleus and tidal tails), M5 (Merger remnant, diffuse merger remnant without bright tidal tails). Refer to Figure \ref{FIG:Morph_fig} for examples. Using the new visual classifications, we can compare our results to previous data and models to get a more complete understanding of the merging process.

\subsection{Completeness corrections for merger type versus infrared luminosity}

\begin{figure}[htb]
		\includegraphics[width= 0.45\textwidth]{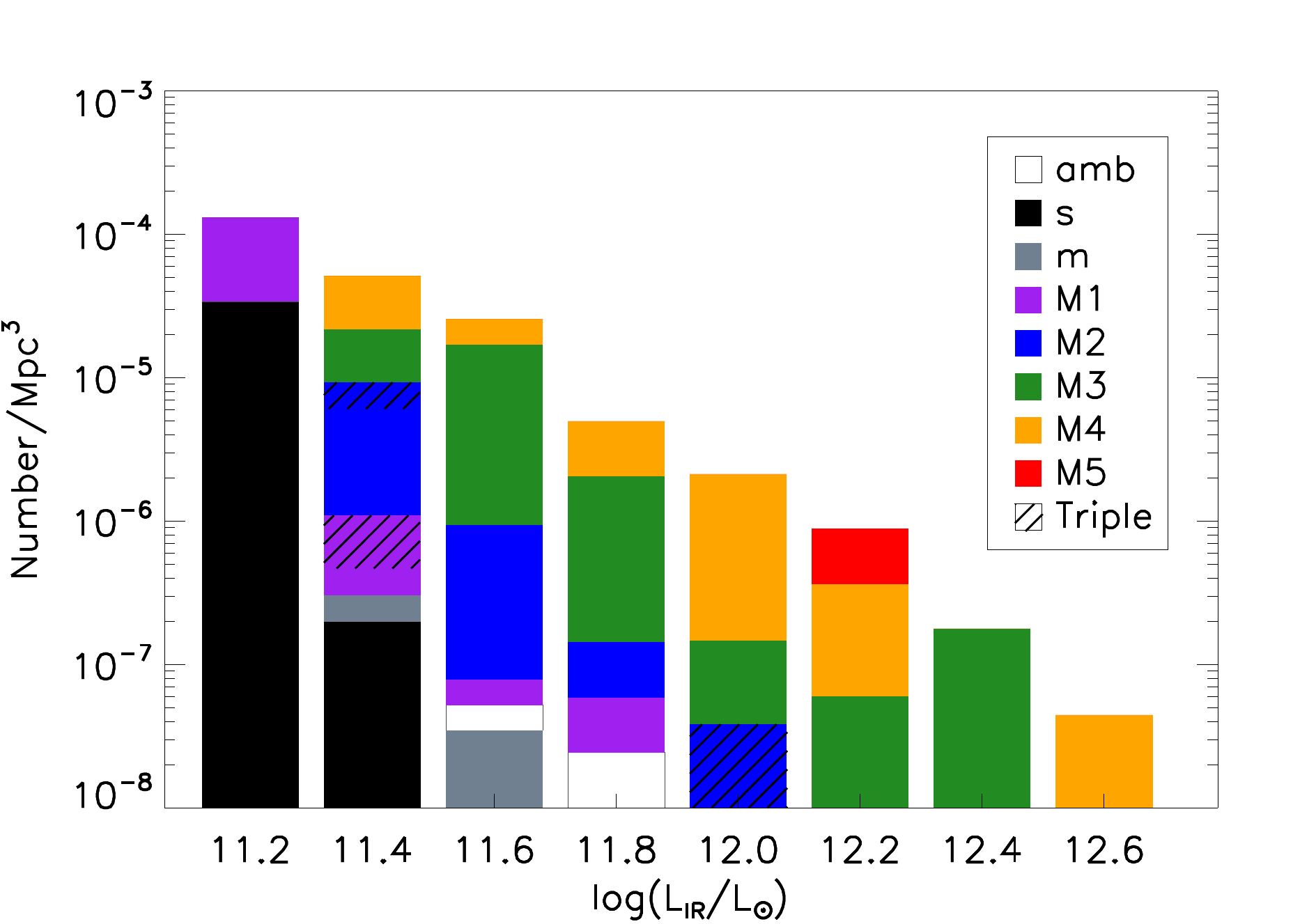}
		\caption{The expected mean number density of objects per Mpc$^3$ in $\Delta L_{\rm IR} = 0.2$ dex bins in a volume corresponding to the volume probed by the largest infrared luminosity bin. The colored bars represent the percentage of galaxies at each interaction stage in each luminosity bin where the total height of each bar represents 100\%. }
		\label{FIG:mstage_VolMpc_percent}
	\end{figure}

Our sample of LIRGs is volume limited and incomplete at the lowest luminosities ($L_{IR} < 10^{11.2} L_\odot$ ).
 We produced a complete volume corrected sample of the number of objects per Mpc in each luminosity bin by comparing our Northern LIRG sample to the full Revised Bright Galaxy Sample and scaling the expected number of objects per bin to the volume probed by the highest luminosity bin, see Figure \ref{FIG:mstage_VolMpc_percent}. 
Since the Northern LIRG sample is not all-sky, all luminosity bins required some completeness correction.
To understand the approximate contributions of each interaction class versus infrared luminosity, the percentage of galaxies present in each class was scaled to the volume corrected sample.
The volume corrected sample allows for a comparison of the relative number of expected galaxies at each interaction stage versus infrared luminosity.

The fraction of interacting galaxies clearly increases with infrared luminosity and at infrared luminosities $L_{IR} <  10^{11.4} L_\odot$ non-interacting galaxies dominate to volume.
All galaxies above an infrared luminosity of $10^{11.5} L_\odot$ are interacting systems while only  $60\%$ of galaxies with infrared luminosities between $10^{11.3}$ to $10^{11.5} L_\odot$ are interacting. The fraction of interacting galaxies from $10^{11.1}$ to $10^{11.3} L_\odot$ may be as low as $14\%$ however this luminosity bin was not fully sampled. 
 Only $54\%$ of the galaxies with infrared luminosities between $10^{11.1}$ to $10^{11.9} L_\odot$ are major mergers when the total number of galaxies in the volume is considered. Therefore, it is the lowest luminosity bin from  $10^{11.1}$ to $10^{11.3} L_\odot$ that adds a large number of non-interacting galaxies to the LIRG population and causes the LIRGs to be a mix of galaxy types from non-interacting to major merging systems.

\subsection{Merger timeline}

\begin{figure}[h!]
	\includegraphics[width= 0.49\textwidth]{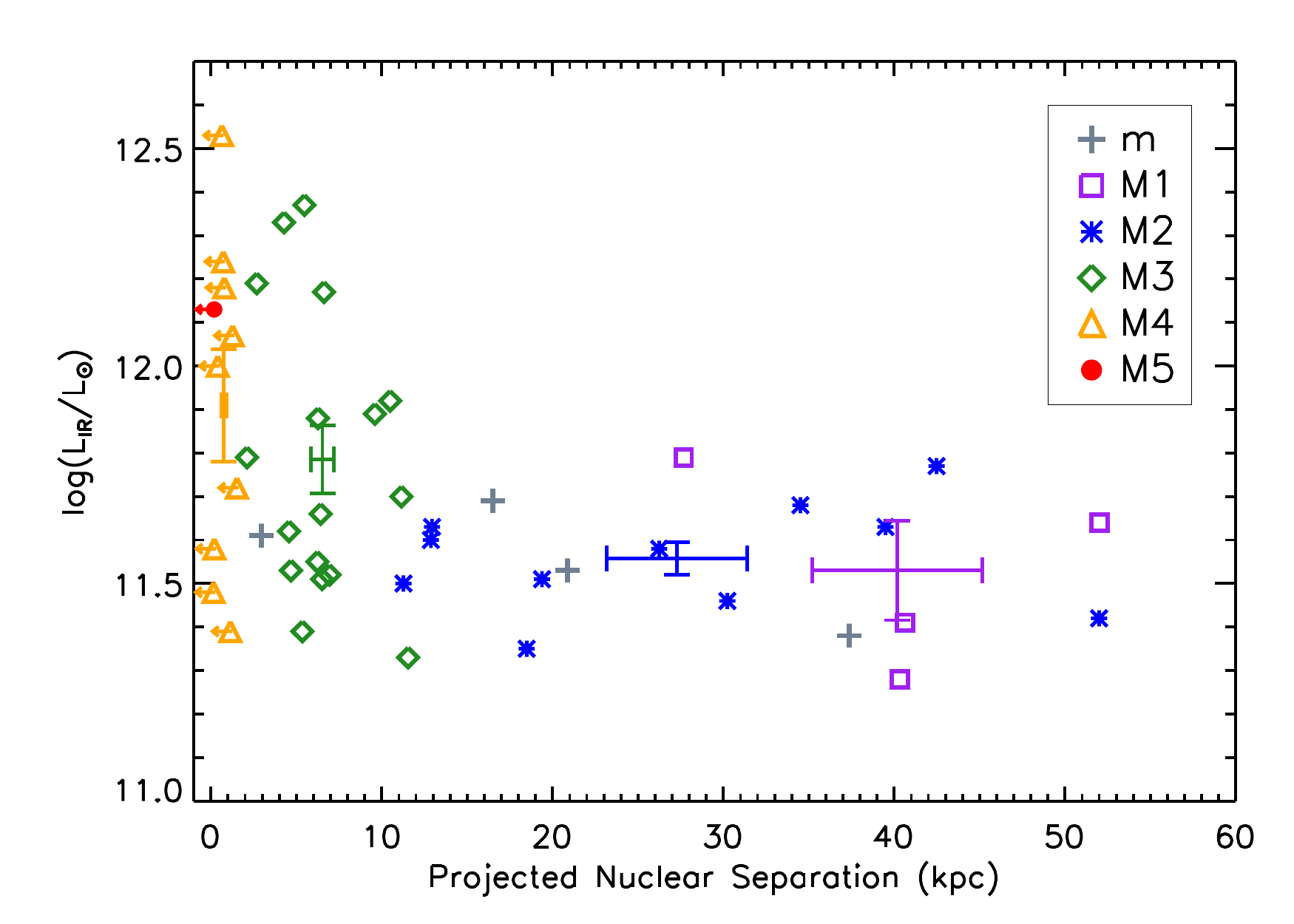}
	\caption{Galaxy infrared luminosity versus projected nuclear separation. Individual objects are color coded according to their merger stage.  The minimum measurable nuclear separation (typically $0.2 - 1.5$\,kpc) is determined by the FWHM of the nucleus as shown by the placement of the M4 and M5 objects with arrows representing upper limits. Error bars correspond to the mean nuclear separation, infrared luminosity, and error on the mean at each major merger stage. }
	\label{FIG:nsep}
\end{figure}

Projected nuclear separation is an easily determined property of interacting galaxies and often used as an indication of merger stage. We compare our visual classification and infrared luminosity to the projected nuclear separations in Figure \ref{FIG:nsep}. For galaxies with only one visible nuclei, the minimum measurable nuclear separation ($\sim0.2-1.5$\,kpc) is determined to be the FWHM of the nucleus and considered an upper limit. We chose to use the nuclear FWHM, instead of just the seeing limit, since the nuclei are imbedded in extended emission of the galaxy and may not be perfect point sources themselves. 
Galaxies with projected nuclear separations less than 2\,kpc are by definition either M4 or M5 since that is larger than the maximum nuclear separation limit set by our observations. 

Detailed dynamical models of galaxy interactions can provide merger timescales for observations. The modeling code Identikit \citep{Barnes:2009aa} uses N-body simulations to explore the dynamical parameter space and determine the best fit model for a merging system. The total time from first pericenter passage (M2) until the merging of the system (M4) ranges from $\sim$250 to 1200\,Myr and varies with the initial mass of the galaxies \citep{Privon:2013aa}. Once interacting galaxies reach second pericenter passage, the time to coalescence of the nuclei coincides with the free fall timescale and is proportional to the nuclear separation of the galaxies \citep{Barnes:2001aa}. The visual classification of M3 corresponds to the time from second pericenter passage until the merging of the nuclei when the galaxies enter merger stage M4. Galaxies spend 75\% to 85\% of the time from the initial interaction until merger stage M4. The merger stage M4 lasts as long as it takes the tidal tails to fall in or fade.
	
While it is clear that the projected nuclear separation is a useful tool, it cannot be used to completely distinguish between all merger stages since projection effects can cause galaxies to appear closer than they actually are. There is a clear trend where earlier merger stages have larger projected nuclear separations than later merger stages.
All major mergers with nuclear separations larger than 15\,kpc are separated or interacting pairs (M1 and M2). However, projected nuclear separation does not distinguish between the separated pairs and interacting pairs which have already experienced first passage.
 Galaxies with nuclear separations between 2 to 15\,kpc are a mixture of merging systems with two nuclei and separated interacting pairs (M3 and M2) with 75\% of the galaxies in this range having a merger stage of M3. 

While late-stage mergers span a large range of infrared luminosities from $10^{11.3}$ to $10^{12.6} L_\odot$, earlier stage interactions (M1 and M2) all have infrared luminosities less than $10^{12.0} L_\odot $.  The early merger stages mostly have projected nuclear separations greater than 15\,kpc and all occupy a rather narrow infrared luminosity range from $10^{11.2}$ to $10^{11.8} L_\odot$, see Figure \ref{FIG:nsep}. The fraction of M1 galaxies is fairly consistent in this infrared luminosity range but the contribution of M2 galaxies peaks at $10^{11.5} L_\odot$ contributing to 50\% of the galaxies in that bin and quickly falls to 0\% by $L_{\rm IR}>10^{12.0} L_\odot$.
Although the infrared luminosity of M1 and M2 galaxies are elevated from the normal galaxy population, they show no sign of extreme starbursts ($L_{\rm IR} > 10^{12} L_\odot$). 
It is not until M3 and nuclear separations of less than 15\,kpc that high infrared luminosities of $L_{\rm IR} > 10^{12} L_\odot$ are seen. 
Therefore, the ULIRG stage in local galaxies is not reached until approximately the last 20\% ($< 200$\,Myr) of the merging process.
This is consistent with galaxy interaction models that show elevated star formation after first pericenter passage and a strong burst in star formation at coalescence \citep{Barnes:2004aa,Hopkins:2013aa}

Four M4 galaxies (UGC 08387, NGC 2623, NGC 0034, and UGC 02238) have infrared luminosities $< 10^{12.0} L_{\odot}$ and are also the lowest mass at this merger stage with $M_{*} < 10^{10.6}M_{\odot}$.  It is possible that these galaxies either did not contain enough initial mass to reach the ULIRG stage or they already had a short lived ULIRG stage which used up the fuel for star formation and allowed the infrared luminosity of the galaxies to decrease to less than $10^{12.0} L_\odot$. Future dynamical models including gas mass and star formation could provide the information needed to better understand what might be happening in these ``low stellar mass" M4 galaxies, as well as to determine if total stellar mass for major merger LIRGs might be expected to correlate with $L_{\rm IR}$ in other major merger stages (see discussion below). 

\subsection{Molecular Gas Fraction vs. Merger Stage}

\begin{figure}
	\includegraphics[width= 0.49\textwidth]{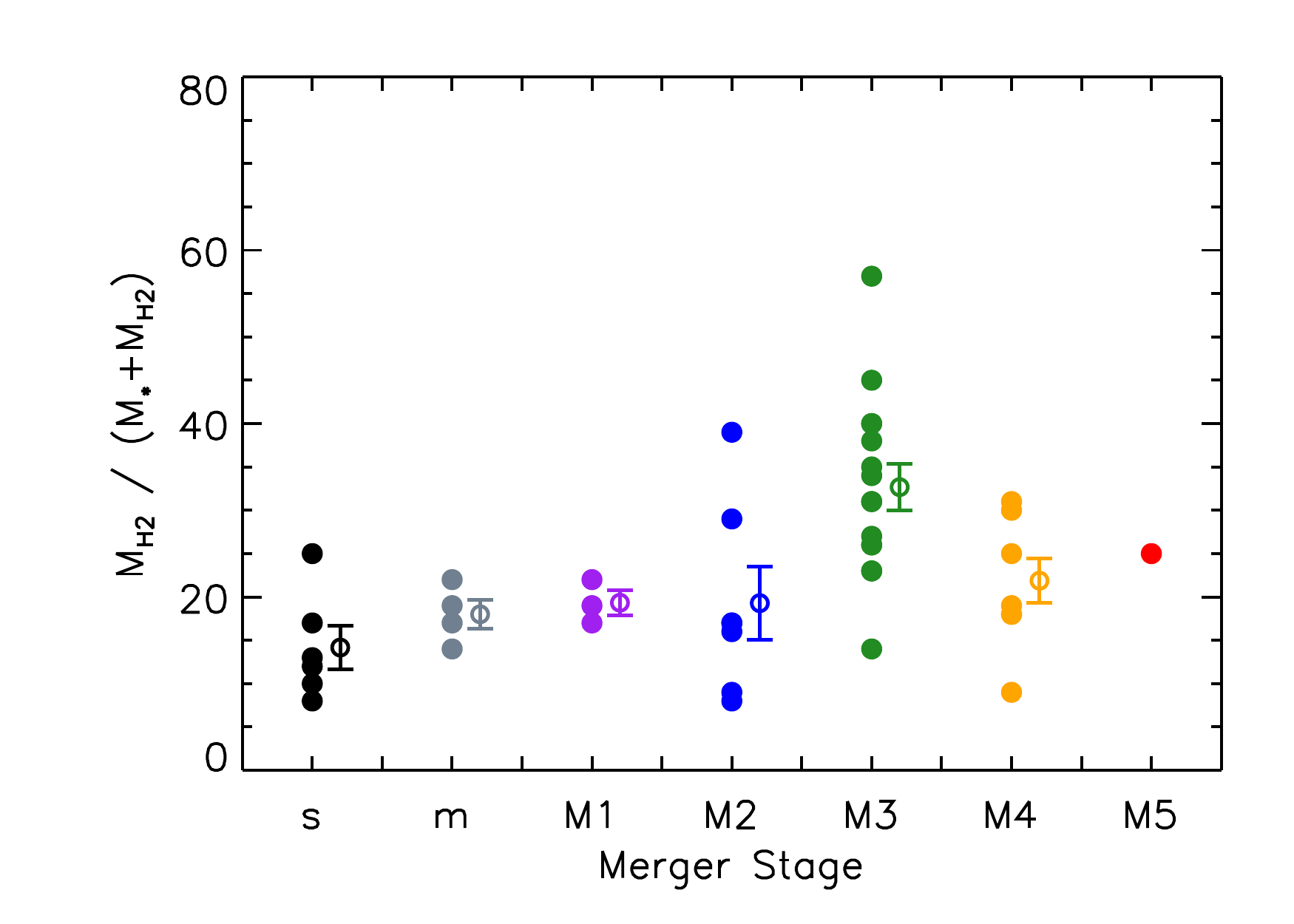}
	\caption{Molecular gas fraction (MGF),  $M_{\rm H_2} / (M_* + M_{\rm H_2})$, is given as a percentage versus merger stage. Empty circle points and error bars correspond to the mean MGF and error of the mean at each stage.}
	\label{FIG:GasMstage}
\end{figure}

The results shown previously in Figure \ref{FIG:LirGas} imply that the mean MGF of LIRGs increases by $\sim \times$2 as sources increase in luminosity above $L_{\rm IR} > 10^{11.5} L_\odot$.   It is also clear from the results shown previously in Figure \ref{FIG:mstage_hist} and Figure \ref{FIG:mstage_VolMpc_percent} that $L_{\rm IR} > 10^{11.5} L_\odot$ represents a luminosity threshold above which gas-rich major mergers become the dominant fraction of extragalactic systems.   This raises an obvious question as to whether the MGF might be related to merger stage.  To explore this hypothesis we plot the MGF versus merger stage in Figure \ref{FIG:GasMstage}.  The results shown in Figure \ref{FIG:GasMstage} imply an increased dispersion in the MGF starting with merger stage M2, with a clear increase in mean MGF to 33\% associated with stage M3, before declining to 22\% in stage M4/5. 

 Table \ref{TAB:Avg_properties} gives a summary of the mean MGF versus visual classification type, showing that the mean MGF increases by $\sim$70\% from merger stage M1 to stage M3 before declining in stage M4 to a value just above that associated with stage M2.  Table \ref{TAB:Avg_properties} also presents the mean total stellar mass of galaxies versus merger stage.  One interesting result is that we find no evidence for a significant increase in mean $M_*$  between major merger stages M1 and M4, although given our relatively small sample size and the corresponding uncertainty in mean $M_*$  ($\sim 0.2$ dex), it is not possible to rule out an increase in $M_*$  as large as $10^{10} M_{\odot}$ between stages M1 and M4.  However, what is clear from the values listed in Table \ref{TAB:Avg_properties}, is that single (s) objects indeed have a value of  $M_*$ that is approximately half that of the major mergers, and minor mergers (m) have a mass in between, as might be expected from the definitions of each class.   

It is also interesting to note that the increase of $\sim$0.3\,dex in the mean MGF between stages M1 and M3 is similar to the increase in mean $L_{\rm IR}$ between these stages, suggesting that it is simply the increased supply of molecular gas that is fueling an enhancement in star formation, although one cannot immediately rule out contributions to $L_{\rm IR}$ from other sources, most notably AGN, without first obtaining high resolution maps of both the molecular gas and infrared luminosity.  However, as shown by \citet{Yuan:2010aa} and \citet{Iwasawa:2011aa}, the majority of GOALS objects hosting powerful AGN are found in stages M4 and M5 where Figure \ref{FIG:GasMstage} shows a decline in the mean MGF, suggesting that even when AGN may contribute a substantial fraction of the observed infrared luminosity they may simultaneously act to decrease the total molecular gas content, e.g. via dissociation of H$_2$ and/or expulsion of molecular gas from the host galaxy by powerful, AGN-driven winds \citep[e.g.][]{Fischer:2010aa, Sturm:2011aa, Veilleux:2013aa, Genzel:2014aa, Tombesi:2015aa}.    

\begin{table}[htb]
	\caption[]{Mean properties of Galaxies vs. Merger Stage}
	\label{TAB:Avg_properties}	
	\centering
	\begin{tabular}{l c c c c }
\hline
\hline \\

	   Type	& log($L_{\rm IR}$) 	& \bf{log($M$)}  	&   $N_{\rm sep}$	& MGF 	\\
	    		&  ($L_{\odot}$)		& \bf{($M_{*}$)}	&    (kpc)	 		&  (\%)  	 \\

\hline \\
 s	&$ 11.29\pm0.03$	& $10.59\pm0.10$	&  	------ 			& $ 14.2\pm2.5$	\\
m	&$ 11.55\pm0.07$	& $10.79\pm0.08$	& $ 19.4\pm7.1$	& $ 18.0\pm1.7$	\\	
M1	&$ 11.53\pm0.11$	& $10.88\pm0.04$	& $ 40.2\pm5.0$	& $ 19.3\pm1.5$	\\
M2	&$ 11.56\pm0.04$	& $10.86\pm0.08$	& $ 27.2\pm4.1$	& $ 19.3\pm4.2$	\\
M3	&$ 11.79\pm0.08$	& $10.71\pm0.06$	& $ 6.52\pm0.67$	& $ 32.7\pm2.7$	\\
 M4	&$ 11.91\pm0.13$	& $10.79\pm0.11$	& $ < 0.76	\pm0.16$	& $ 21.4\pm2.9$	\\
 M5	&$ 12.13$			& $ 10.91$			&$  < 0.19$		&	 25		 \\
\hline
\vspace{0.1truein}
\end{tabular}
\end{table}


Before discussing the origin of the implied increase in the MGF during major gas-rich mergers it is prudent to first consider the possibility that the results shown in Figure \ref{FIG:LirGas} do not actually represent an increase in the MGF, but instead may be due to an error in the determination of $M_{\rm H_2}$, which has been calculated from the product of the observed CO(1-0) luminosity, $L_{\rm CO}$, and the conversion factor, $X_{\rm CO} $.   Given that all of our objects were observed with a beam much larger than the galaxy optical diameter, and with sufficient sensitivity such that the reported errors in $L_{\rm CO}$ are typically $< 15$\%, and that there is {\it no evidence} that the any uncertainty in total $L_{\rm CO}$ is correlated with merger stage, it is highly unlikely that uncertainty in $L_{\rm CO}$ has anything to do with the behavior of $M_{H_2}$ vs. merger stage as observed in Figure \ref{FIG:LirGas}.   Thus uncertainty in $X_{\rm CO} $ is the only issue left to consider.

 Our adopted value for the CO$\rightarrow$H$_2$ conversion factor, $X_{\rm CO}  = 3.0 \times 10^{20}$\,H$_2$\,cm$^{-2}$(K\,km\,s$^{-1}$)$^{-1}$, is based on our previous observations of very large samples of resolved molecular clouds in the Milky Way \citep{Scoville:1987aa, Solomon:1987aa}.  Others have argued for both higher and lower values for the conversion factor, typically in the range $2 \times 10^{20}$ \citep{Bolatto:2013aa}, and $4 \times 10^{20}$ \citep{Draine:2007aa}.  A much lower value of $\sim 0.8 \times 10^{20}$ \citep{Solomon:1997aa, Downes:1998aa} is often adopted for ULIRGs, but this low value was based on critical assumptions about the size and temperature of the compact, nuclear CO emission regions in ULIRGs, which have since proven to be incorrect.  A more detailed analysis of the molecular gas content of galaxies at both low and high redshift and infrared luminosity is given in a recent paper by \citet{Scoville:2016aa}; this paper includes a detailed discussion of the different methods used to derive the $X_{\rm CO}$ and concludes that a value of $3 \times 10^{20}$ provides the most accurate estimate of the mass of H$_2$ gas.   

The effect of adopting a $single$ different value (lower or higher) for the conversion factor simply results in a corresponding rescaling of our computed values for the MGF, and thus our observed trends of MGF versus morphology classification, and in particular for major merger stage, would be unaffected.   Even if we were to assume that the much lower value of the conversion factor originally postulated by \citet{Solomon:1997aa} for ULIRGs should have been adopted for the ULIRGs in our sample, we note that nearly all of our ULIRGs are found in stages M4 and M5, which would then imply lower values for the median MGF only in these stages, thus enhancing the drop in MGF following stage M3 while leaving the rise in MGF observed in stages M2 - M3 unaffected.  

Finally, we note that the extremely compact nuclear concentrations of molecular gas found in many ULIRGs are not typical of what is observed in LIRGs in stages M2 - M3.   
Observations of objects at these merger stages typically show that the majority of the molecular gas is still distributed on large scales (2--10\,kpc), as found for the M2 galaxy ``The Antennae" \citep[Arp244: e.g.][]{Gao:2001aa, Zhu:2003aa}\footnote{ \citet{Zhu:2003aa} used multi-transition CO data from different telescopes, along with 850$\mu$m continuum observations, to derive a value of $X_{\rm CO}$ for ``The Antennae" that was as much as $\times 10$ lower than the Milky Way value.  However, \citet{Scoville:2016aa} used more accurate $Herschel$-SPIRE maps, along with better calibrated CO(1-0) maps, to recalculate a conversion factor for ``The Antennae" that is consistent with the value of $X_{\rm CO}$ adopted in this paper.} and the M3 galaxy NGC 5256 \citep[Arp266:][]{Mazzarella:2012aa}.   Such strong nuclear concentrations of molecular gas are only expected to occur near the final stages of the merger process (stages M4-M5) when the nuclei have coalesced, and even then, simulations suggest that the fraction of molecular gas that is found in the inner kiloparsec can be as low as 30-40\% \citep{Barnes:1996aa, Barnes:2002aa} depending on initial conditions (merger geometry, relative rotation axes, etc.).    

The observed correlation between high MGF and high $L_{\rm IR}$ shown in Figure 4, as well as the enhancement in the MGF during stages M2-M3 of major mergers shown in Figure 7, are important new results, but could these findings possibly be due to selection effects, for example by somehow excluding LIRGs with lower MGF?   We consider this possibility to be unlikely given that the GOALS selection criteria has no a-priori knowledge of $M({\rm H_2})$ and/or merger stage.  A simpler explanation is that all galaxy systems with $L_{\rm IR} > 10^{11}  L_\odot$ have enhanced reservoirs of molecular gas, and that major mergers serve to further enhance the reservoir of molecular gas. We suggest that the most likely mechanism for an enhancement in the MGF is the conversion of a preexisting large reservoir of atomic gas (HI), initially at large galactocentric radii, into molecular gas (H$_2$) as it is drawn into the central regions of the individual galaxies as the merger progresses.  Strong evidence for this process already exists from numerical simulations of gas-rich major mergers \citep[e.g.][]{Barnes:2002aa}.  Such a process is also suggested by recent observations which show a ``flattening" in the metallically gradient during gas-rich major mergers \citep{Kewley:2010aa,Rich:2012aa} that are consistent with a substantial amount of low metallicity HI gas being added to a pre-existing, high metallicity central gas supply. Finally, direct observational evidence for the conversion of HI to H$_2$ during major mergers has already been presented by \citep{Mirabel:1989aa} who used their Arecibo HI survey of LIRGs to show a clear increase in the H$_2$/HI ratio versus $L_{\rm IR}$ and merger stage.

\vspace{0.3truein}

\section{Conclusions}
\label{SEC:Conclusions}

 We have carried out a new analysis of the morphology and molecular gas fraction of a complete sample of  65 LIRGs from the GOALS sample, chosen to be visible from the northern hemisphere.  This Northern LIRG Sample spans the full range of infrared luminosities, $L_{\rm IR} = 10^{11} - 10^{12.6} L_\odot$, and galaxy stellar masses,  $M_* \sim 10^{9.5} - 10^{11.6} M_{\odot}$ observed in the full GOALS sample.  Using HST $I$-band and ground-based $I$-band images from Mauna Kea (UH2.2m) and Haleakala (PS1), we have visually classified all the objects using a simplified classification scheme that includes single galaxies (s), minor mergers (m) and major mergers (M), where the latter class has been subdivided into five merger stages (M1 - M5).   We have also compiled measurements of the total molecular gas masses, $M({\rm H_2}$), converted to a common cosmology and CO ``conversion factor", in order to compute molecular gas fractions for individual galaxies in our sample.  

\vspace{0.1truein}

We find that: 
\begin{enumerate}

	\item The great majority of LIRGs in the GOALS sample can be straightforwardly assigned to one of our galaxy classes (single, minor merger, major merger).  Objects classified as single galaxies have no clear signatures of recent major or minor interaction or merger.   Objects classified as undergoing a minor merger include pairs with mass ratios in the range $\sim$ 4:1 to 10:1.   Major mergers ($<$4:1) were fairly easily distinguished by prominent tidal debris and/or obvious double nuclei.    Five major merger stages were adopted in order to adequately sample the full merger timeline: M1(pre 1st passage pair), M2 (post 1st passage pair) , M3 (overlapping disks, double nuclei and visible tidal tails), M4 (single nucleus with obvious tidal tails), and M5 (diffuse merger remnant).  

	\item Above $L_{\rm IR} = 10^{11.5} L_\odot$ $all$ objects are mergers, with late-stage major mergers representing $>$90\% of objects with $L_{\rm IR} = 10^{12} L_\odot$.  Below $L_{\rm IR} = 10^{11.5} L_\odot$, single galaxies rapidly become the dominant class, representing $\sim$40\% of the LIRGs at $L_{\rm IR} = 10^{11.3} L_\odot$ and $>$80\% of LIRGs at $L_{\rm IR} = 10^{11.1} L_\odot$.

	\item Early stage major mergers (M1 and M2) represent the largest fraction ($\sim$70\%) of the total merger timescale ($<$$t_{\rm mer}$$>$$\sim 1$ Gyr), but exhibit a fairly narrow range ($\sim$0.5 dex) of infrared luminosity,  ($L_{\rm IR} = 10^{11.3-11.8} L_\odot$).  It is not until stage M3 when the galaxies have strongly overlapped into a single disturbed host that we see an increase in infrared luminosity above $10^{12} L_\odot$.

	\item The molecular gas fraction (MGF) clearly increases during the merging process.  Non-interacting LIRGs have a mean molecular gas fraction of $\sim 14 \pm 2.5$\,\% which increases to $\sim 20 \pm 3.3$\,\% in stage M2, and to $\sim 33 \pm 2.7$\,\% in stage M3 before decreasing to $\sim 22 \pm 2.6$\,\% in stage M4. We attribute the observed rise in the MGF to the conversion of HI to H$_2$ as atomic gas from large radii is swept in to the central regions during the merger process.  The subsequent decrease of the MGF in stage M4 can be attributed to gas consumption from starburst activity and AGN growth as well as ionization due to strong feedback from stellar winds and powerful AGN outflows. 

\end{enumerate}

\section{Acknowledgements}
D.S. acknowledges the hospitality of the Aspen Center for Physics, which is supported by the National Science Foundation Grant No. PHY-1066293.   D.S. and K.L. also acknowledge the Distinguished Visitor Program at the Research School for Astronomy and Astrophysics, Australian National University for their generous support while they were in residence at the Mount Stromlo Observatory, Weston Creek, NSW. 
K.L., D.S. and J. B. gratefully acknowledge funding support from  NASA grant NNX11AB02G.  
V.U wishes to acknowledge partial funding support from the Thirty Meter Telescope International Observatory and the UC Chancellor's Postdoctoral Fellowship Program. ASE and DCK and GCP were supported by NASA grant HST-GO10592.01-A and by the NSF grant AST 1109475.
G.C.P. acknowledges support from a FONDECYT Postdoctoral Fellowship (No. 3150361).
A.S.E. also was supported by the Taiwan, R.O.C. Ministry of Science and Technology grant MoST 102-2119-M-001-MY3.
V.U and C.I. extend appreciation toward the UH TAC for their generous support of this project in awarding telescope time on Maunakea, as well as Colin Aspin and the UH 2.2m Telescope staff for their help and support in the acquisition of the ground-based imaging and optical photometry. 
This research has made use of the NASA/IPAC Extragalactic Database (NED) and the IPAC Science Archive, which are operated by the Jet Propulsion Laboratory, California Institute of Technology, under contract with the National Aeronautics and Space Administration. 
 
  The HST-GOALS survey includes observations taken with the NASA/ESA $Hubble\ Space\ Telescope$, obtained at the Space Telescope Science Institute, which is operated by AURA Inc, under NASA contract NAS 5-2655.  
 The Pan-STARRS1 Surveys (PS1) have been made possible through contributions of the Institute for Astronomy, the University of Hawaii, the Pan-STARRS Project Office, the Max-Planck Society and its participating institutes, the Max Planck Institute for Astronomy, Heidelberg and the Max Planck Institute for Extraterrestrial Physics, Garching, The Johns Hopkins University, Durham University, the University of Edinburgh, Queen's University Belfast, the Harvard-Smithsonian Center for Astrophysics, the Las Cumbres Observatory Global Telescope Network Incorporated, the National Central University of Taiwan, the Space Telescope Science Institute, the National Aeronautics and Space Administration under Grant No. NNX08AR22G issued through the Planetary Science Division of the NASA Science Mission Directorate, the National Science Foundation under Grant No. AST-1238877, the University of Maryland, and Eotvos Lorand University (ELTE), and the Los Alamos National Laboratory.


\appendix
\section{Visual Classifications}
Table \ref{TAB:SchemesK13} and  \ref{TAB:SchemesSt13} directly compare our classification (L16) to that of previous classifications of the GOALS objects given in Stierwalt et al.(2013: St13), Kim et al. (2013: K13). 
Some stages, like our distinction of minor mergers (m), have no comparable classification in the other schemes. Kim et al. (2013: K13) also included a merger stage defined as a ``common envelope''  (stage 3) which has no corresponding stage in our classification scheme.
Even though the definition of a class might be the same between the schemes, a galaxy classified as an M2 in our scheme may not have the corresponding classification of $b$ in Stierwalt et al. (2013: St13) since St13 used lower resolution data to classify the galaxies.  
 Table \ref{TAB:SchemesK13} and  \ref{TAB:SchemesSt13}  provide a grid comparison of our classification (L16) to K13 and St13 and give the number of objects in each cell. We color code cells as green where the corresponding definitions of merger stage agree. Classifications that are shifted by only one class to a slightly earlier or later stage in our classification scheme are color coded as yellow and considered a slight change from the previous classification. Major changes are those that required a change of classification by more than a single adjacent merger stage and are colored as orange in the comparison grids.
 
 Fourty-one of the galaxies in our sample were also previously classified by \citet{Kim:2013aa} as seen in table \ref{TAB:SchemesK13}. Our classifications agree fairly well with that of K13 with two objects now classified as ambiguous and 12/41 (29\%) objects requiring a slight change in classification in our scheme.  Half of the slight changes result from K13 inclusion of the intermediate ``common envelope'' (stage 3) class. We classify all of these ``common envelope'' galaxies as M3, double nuclei systems. Only 3/41 (7\%) objects, required a major change from K13 classification. All of the major changes result from our inclusion of minor mergers in the overall classification scheme. 

%
%


\begin{table}[htb]
	\caption[]{Comparison of this work (L16) to K13}
	\label{TAB:SchemesK13}	
	\centering
	\begin{tabular}{| c |c|c|c|c|c|c|c|}
\hline
\diagbox[dir=NW, width=5em]{L16}{K13}& \makebox[2em]{none}& \makebox[2em]{1}&\makebox[2em]{ 2}	
& \makebox[2em]{3}	& \makebox[2em]{4}	& \makebox[2em]{5}	& \makebox[2em]{6}	\\

\hline
		s	& \cellcolor{green}\textbf{0} & 0		& 0		& 0		& 0		& 0	& 0	\\
		\hline	
		m	&  \cellcolor{orange}2		& 0		& 0		& 0		& 0		& \cellcolor{orange}1	& 0	\\
		\hline	
		M1	& 0		&\cellcolor{green}\textbf{3}& 0		& 0		& 0		& 0	& 0	\\
		\hline
		M2	& 0		& \cellcolor{yellow}1		&\cellcolor{green}\textbf{9}	& 0		& 0		& 0	& 0	\\
		\hline
		M3	& 0		& 0		&  \cellcolor{yellow}2		& \cellcolor{yellow}5		&\cellcolor{green}\textbf{5}	&  \cellcolor{yellow}2	& \cellcolor{yellow}1	\\
		\hline
		M4	& 0		& 0		& 0		& 0		&  0		& \cellcolor{green}\textbf{7}&\cellcolor{green}\textbf{0}	\\
		\hline
		M5	& 0		& 0		& 0		& 0		& 0		& 0	& \cellcolor{yellow}1	\\
		\hline
		amb	& 2		& 0		& 0		& 0		& 0		& 0	& 0	\\

\hline
\end{tabular}

\end{table}

\citet{Stierwalt:2013aa} classified 63 of the objects in our sample using IRAC data as seen in table \ref{TAB:SchemesSt13}. Most of the differences in our classifications are the result of St13 using lower resolution and shallower data to perform the classifications. This is most apparent in single nucleus late-stage mergers (St13 stage $d$) where in half of the galaxies we were able to resolve two galaxy nuclei with higher resolution data and re-classify them to earlier merger stages. In total, 22\% of all the objects required only a slight change in merger stage while 16\% had a major change in classification.

\begin{table}[htb]
	\caption[]{Comparison of this work (L16) to St13}
	\label{TAB:SchemesSt13}	
	\centering
	\begin{tabular}{| c |c|c|c|c|c|c|c|}
\hline
 \diagbox[width=5em]{L15}{St13} &  \makebox[3em]{N}& \makebox[3em]{a}&  \makebox[3em]{b}
 & \makebox[3em]{c}	& \makebox[3em]{d}	\\

\hline
		s	& \cellcolor{green}\textbf{10} & \cellcolor{orange}1	& 0		& 0		& \cellcolor{orange}1	\\
		\hline
		m	& \cellcolor{orange}2		& \cellcolor{orange}1		& 0		& 0		& \cellcolor{orange}1	\\
		\hline	
		M1	& 0		& \cellcolor{green}\textbf{5}& \cellcolor{yellow}3		& 0		& 0	\\
		\hline
		M2	& 0		& \cellcolor{yellow}1		& \cellcolor{green}\textbf{7}& 0		& \cellcolor{orange}2 \\
		\hline
		M3	& 0		& \cellcolor{orange}1		& \cellcolor{yellow}4		& \cellcolor{green}\textbf{5}& \cellcolor{yellow}7	\\
		\hline
		M4	& 0		& 0		& 0		& 0		& \cellcolor{green}\textbf{9}\\
		\hline
		M5	& 0		& 0		& 0		& 0		& \cellcolor{green}\textbf{1}\\
		\hline
		amb	& 1		& 1		& 0		& 0		& 0	\\

\hline
\end{tabular}

\vspace{0.1truein}
\end{table}

We provide further details and justifications for the classifications of all the galaxies.  For each galaxy we describe the visual evidence (e.g. tidal tails, loops, number of nuclei) used for the classification, as well as the corresponding $\Delta v$ and nuclear separations of galaxies in the system. Major differences in classifications between our scheme and previous classifications of St13 and K13 are also addressed. All galaxies in our sample are listed by RA order below.

\vspace{0.2truein} 

\underbar{NGC0034} [$L_{\rm IR} = 10^{11.48} L_\odot$] Classified as {\bf M4} based on the observed single nucleus and prominent long tidal tail ($\sim$30\,kpc) extending to the NE and second tidal  ``loop" to the NW. \citet{Schweizer:2007aa} previously classified this object as a single nucleus merger remnant that resulted from a major merger with an estimate mass ratio of $\sim 1/3 <$ m/M $<  2/3$.  

\underbar{MCG-02-01-051} (= Arp256)  [$L_{\rm IR} = 10^{11.46} L_\odot$]  Classified as {\bf M2} based on the observed wide separated pair where the northern and southern galaxies have prominent tidal arms and/or plumes. The two galaxies Arp 256-01 and Arp 256-02 have a nuclear separation of 30\,kpc and a $\Delta v$ of 68\,km/s. 
\citet{Chien:2010aa} found that MCG-02-01-051 is best modeled as an early-stage interaction, which agrees with our classification of {\bf M2}.

\underbar{NGC0232}  [$L_{\rm IR} = 10^{11.43} L_\odot$]  has been previously classified as a non-interacting system, or as simply a member of a Compact Group. We classify this object as {\bf M2} in a triple system, with NGC0232 being the most infrared luminous galaxy in the system. NGC0232 is interacting with NGC0235 which is $\sim$53\,kpc to the NE, with a $\Delta v$ of 123\,km/s, and a faint tidal bridge connecting the two galaxies. NGC0235 has two nuclei and is classified as a minor interaction. 
We therefore classify NGC0232 as being part of a {\bf TM2} system.

\underbar{IC1623}  (= VV114)  [$L_{\rm IR} = 10^{11.66} L_\odot$]  Classified as {\bf M3} based on clearly overlapping -- one edge-on (E) and one face-on (W) disk -- with projected nuclear separation of $\sim$6\,kpc, in addition to a long ($\sim$70\,kpc), moderately faint, curved tidal tail extending  to the N and E. K13 favored a ``common envelope" classification of this system.

\underbar{MCG-03-04-014}   [$L_{\rm IR} = 10^{11.63} L_\odot$]  was previously classified as a non-interacting galaxy by St13. The galaxy has a diffuse disk and a possible disconnected diffuse tail to the west. It is unclear if the diffuse structure $\sim$37\,kpc to the West is the remnant of a tidal tail and we therefore leave the classification as {\bf amb}.

\underbar{CGCG436-030}  [$L_{\rm IR} = 10^{11.68} L_\odot$]  Classified as {\bf M2} based on non-overlapping (projected nuclear separation $\sim 35$\,kpc), tidally-disturbed disks with obvious tidal tails.  

\underbar{IRASF01364}  [$L_{\rm IR} = 10^{11.79} L_\odot$]  Classified as {\bf M3} based on highly disturbed common disk with small projected nuclear separation ($\sim 2$\,kpc), and obvious tidal tail(s) extending to the WSW.   Nearby foreground bright star (S) may have inhibited previous attempts to properly classify this object.  We note that K13 classified this source as a single nucleus system.

\underbar{IIIZw035}  [$L_{\rm IR} = 10^{11.62} L_\odot$] Classified as {\bf M3} based on overlapping, nearly edge-on disturbed disks with projected nuclear separation of $\sim 5$\,kpc. K13 favored a ``common envelope" classification of this system and St13 classified this system as a separated galaxy pair (stage $a$).

\underbar{NGC0695}  [$L_{\rm IR} = 10^{11.69} L_\odot$]  has previously been classified as a non-interacting galaxy by St13; however there is a minor ($>$4:1) companion $\sim$16\,kpc\,NW of the main galaxy along with the appearance of a tidal perturbation,  which leads us to classify this system as a minor merger ({\bf m}).

\underbar{NGC0958}  [$L_{\rm IR} = 10^{11.22} L_\odot$]  Classified as {\bf s} based on the appearance of a single, large ($\sim 70$\,kpc diameter), relatively edge-on spiral galaxy. Two small ($>$10:1) possible satellites to the E do not appear to be associated with signs of tidal disturbance. 

\underbar{NGC1068}  [$L_{\rm IR} = 10^{11.4} L_\odot$]  Classified as {\bf s}.  This well-known, nearby Seyfert\,2 galaxy does not appear to be currently interacting with another galaxy.  Although relatively large in angular extent, NGC1068 is physically small compared to the more distant objects in our sample.

\underbar{UGC02238}  [$L_{\rm IR} = 10^{11.39} L_\odot$]  Classified as  {\bf M4} based on a single nucleus and the appearance of a highly disturbed disk with two tidal tails to the S and tidal plumes to the N.  

\underbar{UGC02369}   [$L_{\rm IR} = 10^{11.60} L_\odot$]  Classified as {\bf M2} based on the appearance of two highly disturbed disks with projected nuclear separation of $\sim$13\,kpc.

\underbar{IRASF03359}  [$L_{\rm IR} = 10^{11.51} L_\odot$] Classified as {\bf M3} based on overlapping, one edge-on and one face-on disturbed disks with projected nuclear separation of $\sim$7\,kpc.   K13 favored a ``common envelope" classification of this system while St13 classified it as a single nucleus late-stage merger (stage $d$).

\underbar{UGC02982}  [$L_{\rm IR} = 10^{11.20} L_\odot$]  Classified as  {\bf s} based on the appearance of a single, relatively undisturbed, clumpy spiral disk. St13 classified this galaxy as a single nucleus late-stage merger (stage $d$).

\underbar{ESO550-IG025}  [$L_{\rm IR} = 10^{11.50} L_\odot$]  Classified as  {\bf M2} based on the appearance of two clearly disturbed disks with projected nuclear separation of $\sim$11\,kpc.  

\underbar{NGC1614}  [$L_{\rm IR} = 10^{11.61} L_\odot$]  Classified as  {\bf m}. The galaxy has been previously defined as a late-stage major merger (equivalent to our class M4) by both K13 and St13. However, this object has a fairly ordered dominant spiral structure associated with the single bright nucleus, in addition to a putative ``tidal tail" extending to the SW.  The combination of the apparent single merged nucleus and the apparent bright SW tail implies a relatively short post-merger timescale that is inconsistent with the fairly ordered large scale spiral structure of the previously assumed  merged disk \citep{Barnes:2002aa}. It seems much more probable that NGC1614 is the result of a minor merger between a larger face on spiral and a smaller edge on disk, where the mass ratio of the interacting pair is likely $>$ 4:1. The apparent SW tail would then be more correctly interpreted as the smaller disk viewed edge on, and the bright point source at position 68.499 RA, -8.58 DEC would then be the nucleus of the smaller galaxy at a projected nuclear separation of $\sim$3\,kpc. The nucleus of the minor companion was also identified by \citet{Vaisanen:2012aa} and a high mass ratio of $\ge$ 4:1 was previously suggested by \citet{Rothberg:2006aa} and \citet{Xu:2015aa}.

\underbar{IRASF05189}  [$L_{\rm IR} = 10^{12.13} L_\odot$]  Classified as {\bf M5} based on the detection of a single compact, slightly off-center nucleus, a disk featuring ripples and shell-like structures, plus faint tidal tails to the SE and NE.  This object was originally identified as a candidate ``infrared quasar" \citep{Sanders:1988ab}, based on its extreme luminosity and Seyfert\,1 broad-line optical spectrum. 

\underbar{NGC2623}   [$L_{\rm IR} = 10^{11.58} L_\odot$]  Classified as  {\bf M4} based on the detection of a single nucleus embedded in a disturbed host with two large tidal tails extending to the NE and SW. 

\underbar{IRASF08572}   [$L_{\rm IR} = 10^{12.17} L_\odot$]  Classified as {\bf M3} base on the appearance of two highly disturbed, partially overlapping disks with projected nuclear separation of $\sim$7\,kpc. 

\underbar{UGC04881}   [$L_{\rm IR} = 10^{11.70} L_\odot$]   Classified as {\bf M3} based on the appearance of two highly disturbed, partially overlapping disks with a bright tidal tail to the SE and prominent tidal debris to the WNW.  The projected nuclear separation is $\sim$11\,kpc. 

\underbar{UGC05101}   [$L_{\rm IR} = 10^{12.0} L_\odot$]  Classified as  {\bf M4} based on the detection of a single bright nucleus embedded in a disturbed disk with a large tidal tail extending to the W and a long looped tail extending from the NE around to the SW. 

\underbar{MCG+08-18-013}   [$L_{\rm IR} = 10^{11.33} L_\odot$]  Classified as {\bf M1} based on the appearance of two widely separated spiral disks (MCG+08-18-013/012 = CGCG239-011),  which both exhibit slight tidal distortions along a line connecting the two nuclei.  MCG+08-18-013 is the dominant infrared source.  Both disks are approximately equal in mass and have a projected nuclear separation of $\sim$52\,kpc and a $\Delta v$ of 231\,km/s. Given that there is a third smaller galaxy ($>$4:1) to the SE of MCG+08-18-013 that may be interacting with the dominant infrared source, we also consider the possibility that this source is a Triple system where MCG+08-18-013 and its smaller SE ``companion" represent a  minor merger. 

\underbar{IC0563/4}   [$L_{\rm IR} = 10^{11.28} L_\odot$]  Classified as  {\bf M1} based on the appearance of two slightly disturbed disk galaxies with projected nuclear separation of $\sim$40\,kpc and a $\Delta v$ of only 12\,km/s.

\underbar{NGC3110}   [$L_{\rm IR} = 10^{11.41} L_\odot$]  Previously been classified as a non-interacting galaxy. However, NGC 3110 has a companion galaxy (MCG -01-26-013) to the southwest with a projected separation of $\sim$31\,kpc and $\Delta v$ of 235\,km/s.  Therefore we classify NGC3110 is a {\bf M1} major merger.

\underbar{IRASF10173}   [$L_{\rm IR} = 10^{11.79} L_\odot$]  This object appears to have a single nucleus (at our HST resolution) with faint tidal tails extending to the north and south, which would have resulted in an M4 classification.   However, a second smaller disturbed galaxy (SDSS CGB24551.1) can be seen at a projected separation of 28\,kpc to the west, but this possible companion has no reported redshift.  Although we favor an {\bf M4} classification, we have listed this object as ``ambiguous" since we cannot definitively rule out a minor merger (m) classification. St13 favored a classification of ``separated galaxy pair'' (stage $a$).

\underbar{IRASF10565}  [$L_{\rm IR} = 10^{12.05} L_\odot$]   Classified as a major merger {\bf M2} based on the appearance of two disturbed disk galaxies (W and NE) connected by a tidal bridge and a projected nuclear separation of $\sim$23\,kpc. The dominant infrared source (IRASF10565-W) appears to also have a second fainter nucleus $\sim$6\,kpc to the east of the main nucleus. We therefore classify this object as a potential triple system (TM2).

\underbar{MCG+07-23-019}  [$L_{\rm IR} = 10^{11.63} L_\odot$]   Classified as {\bf M2} based on the appearance of two highly disturbed galaxies - one edge on and the other a ``ring system".  The projected nuclear separation is $\sim$13\,kpc. St13 interpreted the ring system as a tidal tail and classified this system as a single nucleus late-stage merger (stage $d$).

\underbar{CGCG011-076}   [$L_{\rm IR} = 10^{11.38} L_\odot$]  Classified as  {\bf m} based on the appearance of a distorted disk with a long tidal feature to the WSW connected to a smaller ($>$4:1) companion with a projected nuclear separation of $\sim$37\,kpc. This system had been previously classified as a separated galaxy pair (stage $a$) by St13.

\underbar{IC2810}    [$L_{\rm IR} = 10^{11.64} L_\odot$]   Classified as  {\bf M1} based on the appearance of two slightly disturbed disks with large projected nuclear separation of $\sim$52\,kpc and a $\Delta v$ of 103\,km/s.

\underbar{NGC3690}  (= Arp299)   [$L_{\rm IR} = 10^{11.89} L_\odot$]   Classified as  {\bf M3} based on the appearance of two highly overlapping disks with clear tidal features.   The projected nuclear separation is $\sim$10\,kpc. K13 classified this as a ``common envelope'' system.

\underbar{IRASF12112}   [$L_{\rm IR} = 10^{12.33} L_\odot$]  Classified as  {\bf M3} based on the appearance of a highly disturbed system with obvious tidal features and two nuclei with projected separation of $\sim$4\,kpc. K13 also classified this galaxy as a double nucleus system while St13 favored a single nucleus late-stage merger classification.

\underbar{IRASF12224}  [$L_{\rm IR} = 10^{11.32} L_\odot$]   Classified as  {\bf s} based on the appearance of a single, nearly face-on barred spiral disk with no obvious signs of tidal interaction. 

\underbar{UGC08058}  (Mrk 231)   [$L_{\rm IR} = 10^{12.53} L_\odot$]   Classified as  {\bf M4} based on the single bright nucleus slightly off-centered with respect to a very large host galaxy, with two large tidal tails extending to the N and S from the E edge of the merged disk.  This object has previously been characterized as a ``infrared quasar" based on its extreme luminosity and a Seyfert\,1 optical spectrum. 

\underbar{NGC4922}   [$L_{\rm IR} = 10^{11.33} L_\odot$]  Classified as  {\bf M3} based on the appearance of two partially overlapping, highly disturbed disks with distinct tidal features.  The projected separation of the two nuclei is $\sim$12\ kpc. 

\underbar{ESO507-G070}   [$L_{\rm IR} = 10^{11.53} L_\odot$]   Classified as  {\bf M3} based on the appearance of two overlapping, slightly edge-on, highly disturbed disks with distinct tidal features.  The projected separation of the two nuclei is $\sim$5\,kpc. St13 only resolved one nucleus and therefore classified the system as a single nucleus, late-stage merger (class $d$). K13 also only identified one nucleus and classified the system as a late-stage merger (stage 6).

\underbar{IC0860}   [$L_{\rm IR} = 10^{11.10} L_\odot$]   Classified as {\bf s} based on lack of any clear sign of an interaction in this relatively compact, slightly edge-on disk. 

\underbar{VV250a}   [$L_{\rm IR} = 10^{11.77} L_\odot$]  Classified as  {\bf M2} based on the appearance of two widely separated, but clearly disturbed disks connected by a tidal ``bridge" and in addition, exhibiting prominent tidal tails to the NW and SE.

\underbar{UGC08387} (= Arp193)   [$L_{\rm IR} = 10^{11.72} L_\odot$]  Classified as  {\bf M4} based on the detection of a single nucleus in a disturbed disk with two equally prominent tidal tails to the SE and SW. 

\underbar{NGC5104}   [$L_{\rm IR} = 10^{11.25} L_\odot$]  Classified as  {\bf s} based on the appearance of a single, nearly edge-on spiral disk with no obvious signs of tidal interaction. 

\underbar{MCG-03-34-064}  [$L_{\rm IR} = 10^{11.19} L_\odot$]   Classified as {\bf s}. The galaxy visible to the north west is MCG-03-34-063 and is in projection. St13 previously classified this galaxy as a close galaxy pair (stage $a$). MCG-03-34-063 has a projected nuclear separation of $\sim$41\,kpc but the large reported relative velocity of the two objects, $\Delta v = 1435$\,km/s, argues against these galaxies being a close pair. MCG-03-34-064 is therefore classified as a single (s), non-interacting galaxy.

\underbar{NGC5135}  [$L_{\rm IR} = 10^{11.29} L_\odot$]    Classified as  {\bf s} based on the appearance of a single, nearly face-on barred spiral galaxy.

\underbar{NGC5256}  (= Mrk266)  [$L_{\rm IR} = 10^{11.52} L_\odot$]   Classified as {\bf M3} based on the appearance two nuclei (projected separation of $\sim$7\,kpc) embedded in the center of a large, highly disturbed system with numerous large tidal features (e.g. loops, bridges and tails). K13 favored the ``common envelope'' classification while St13 classified the system as an interacting galaxy pair (stage $b$).

\underbar{NGC5257/8}    [$L_{\rm IR} = 10^{11.63} L_\odot$]  Classified as {\bf M2} based on the appearance of two well defined, widely separated disks (projected nuclear separation of $\sim$40\,kpc) connected by a prominent tidal bridge and $\Delta v = 41$\,km/s. 

\underbar{UGC08696}  (= Mrk273)   [$L_{\rm IR} = 10^{12.18} L_\odot$]  Classified as {\bf M4} based on the appearance of very large tidal tails extending to the S and NE from a merged main body.  Although two nuclei have been identified in this system \citep[e.g.][]{U:2013aa}, their small angular separation ($<$1\,arcs, corresponding to $<$1\,kpc) leads us to the M4 visual classification. 

\underbar{CGCG247-020}   [$L_{\rm IR} = 10^{11.35} L_\odot$]   Classified as  {\bf s} based on the appearance of a single, small object with no clear sign of an interaction. 

\underbar{IRASF14348}   [$L_{\rm IR} = 10^{12.37} L_\odot$]   Classified as  {\bf M3} based on the the appearance of a large highly disturbed, double nucleus system with prominent tidal tails to the N and SW.  The projected nuclear separation us $\sim$5.5\,kpc. 

\underbar{VV340a}  [$L_{\rm IR} = 10^{11.79} L_\odot$]   Classified as {\bf M1}.  This object is comprised of a face on spiral and an edge on spiral galaxy. The $\Delta v$ of the galaxies is 65\,km/s. Even though the projected nuclear separation is only 27\,kpc and the edges of the galaxy disks appear close in projection, neither galaxy shows any obvious sign of significant tidal disturbance, suggesting that the actual separation is much larger than the measured projection. We therefore favor a classification of M1, galaxy pair prior to first passage, as opposed to that previously given by St13 as an interacting galaxy pair after first passage (stage $b$).

\underbar{CGCG049-057}   [$L_{\rm IR} = 10^{11.33} L_\odot$]   Classified as  {\bf s} based on the appearance of a single, small object with no clear sign of an interaction. 

\underbar{VV705}   [$L_{\rm IR} = 10^{11.88} L_\odot$]  Classified as  {\bf M3} based on the the appearance of a large highly disturbed, double nucleus system with prominent tidal loops/tails extending from the to the NW and SE.  The projected nuclear separation us $\sim$6\,kpc. St13 previously classified this system as an interacting galaxy pair (stage $b$).

\underbar{IRASF15250}   [$L_{\rm IR} = 10^{12.07} L_\odot$]   Classified as  {\bf M4} base on the appearance of a bright single nucleus off-centered in a highly disturbed disk with a prominent tidal ring or looped tails. 

\underbar{UGC09913} (= Arp220)   [$L_{\rm IR} = 10^{12.24} L_\odot$]  Classified as {\bf M4} based on the appearance of a heavily obscured nuclear region embedded in a highly disturbed disk with prominent tidal tails extending to the N and S from the western edge of the disk. Although two nuclei have been identified in this system \citep[e.g.][]{Baan:1987aa, Graham:1990aa}, their small angular separation ($<$1\,arcs, corresponding to $<$1\,kpc) leads us to the M4 visual classification. 

\underbar{NGC6090}    [$L_{\rm IR} = 10^{11.55} L_\odot$]  Classified as  {\bf M3} based on the detection of two nuclei embedded in the center of a disturbed disk with prominent tidal tails extending to the NE and WSW.   The projected nuclear separation is $\sim$6\,kpc. St13 classified this as a single nucleus late-stage merger.

\underbar{CGCG052-037}   [$L_{\rm IR} = 10^{11.45} L_\odot$]   Classified as  {\bf s} based on the appearance of a single, small object with no clear sign of an interaction.

\underbar{NGC6286}   [$L_{\rm IR} = 10^{11.42} L_\odot$]   Classified as  {\bf M2} based on the appearance of two widely separated disks (projected nuclear separation of $\sim$52\,kpc) with clear signs of tidal disturbance and a $\Delta v = 190$\,km/s. 

\underbar{IRASF17132}   [$L_{\rm IR} = 10^{11.92} L_\odot$]   Classified as {\bf M3} based on the appearance of two highly disturbed disks with projected nuclear separation $\sim$10.5\,kpc. K13 previously classified this system as an interacting galaxy pair (stage 2).

\underbar{ESO602-G025}    [$L_{\rm IR} = 10^{11.34} L_\odot$]  Classified as  {\bf s} based on the appearance of a single, small object with no clear sign of an interaction.

\underbar{IRASF22491}   [$L_{\rm IR} = 10^{12.19} L_\odot$]   Classified as {\bf M3} based on the detection of two nuclei embedded in the center of a highly disturbed disk with prominent tidal tails extending to the E and NW.   The projected nuclear separation is $\sim$2.7\,kpc. K13 and St13 classified this as a single nucleus late-stage merger.

\underbar{NGC7469}  (= Arp 298)   [$L_{\rm IR} = 10^{11.58} L_\odot$]   Classified as {\bf M2} based on the appearance of two widely separated disks (projected nuclear separation of $\sim$26\,kpc), where the dominant infrared source corresponds to the well-known Seyfert\,1 nucleus in the larger, nearly face-on disk.  

\underbar{CGCG453-062}    [$L_{\rm IR} = 10^{11.37} L_\odot$]   Classified as  {\bf s} based on the appearance of a single, small object with no clear sign of an interaction.

\underbar{IC5298}    [$L_{\rm IR} = 10^{11.53} L_\odot$]   Classified as  {\bf m} where the dominant infrared source corresponds to the nucleus of the large barred spiral, and the smaller companion ($>$4:1) to the SW (projected nuclear separation $\sim$21\,kpc) is connected by a faint tidal bridge. St13 classified this object as a non-interacting galaxy.

\underbar{NGC7592}   [$L_{\rm IR} = 10^{11.39} L_\odot$]   Classified as {\bf M3} based on the appearance of two highly disturbed disks (projected nuclear separation $\sim$5.4\,kpc) and prominent tidal tails to the N and SW. St13 classified this system as an interacting galaxy pair (stage $b$).

\underbar{NGC7674}    [$L_{\rm IR} = 10^{11.51} L_\odot$]  Classified as  {\bf M2} based on the appearance of a clear double disk system (projected nuclear separation $\sim$19.4\,kpc), with a tidal bridge connecting the smaller galaxy to the NE to the larger, face-on Seyfert\,2 galaxy which is the dominant infrared source. 

\underbar{NGC7771}    [$L_{\rm IR} = 10^{11.35} L_\odot$]   Classified as  {\bf M2} based on the appearance of a clear double disk system (projected nuclear separation $\sim$18.5\,kpc), with a tidal features  connecting the smaller galaxy to the S to the larger, nearly edge-on galaxy which is the dominant infrared source. 

\underbar{Mrk331}    [$L_{\rm IR} = 10^{11.50} L_\odot$]  The dominant infrared source is a nearly face-on spiral galaxy that we have classified as being part of an {\bf M1} system based on the presence of an edge-on spiral galaxy (UGC12812 at a projected nuclear separation of $\sim$41\,kpc and velocity separation of 158\,km/s), which may itself be interacting with a small nearby companion galaxy.  We also identify the entire 3 object system (KPG593) as a possible triple system (TM1).

\bibliographystyle{apj}
\bibliography{GOALS_morph1_astroph.bib}

\begin{thebibliography}{}
\expandafter\ifx\csname natexlab\endcsname\relax\def\natexlab#1{#1}\fi

\bibitem[{{Armus} {et~al.}(2009){Armus}, {Mazzarella}, {Evans}, {Surace},
  {Sanders}, {Iwasawa}, {Frayer}, {Howell}, {Chan}, {Petric}, {Vavilkin},
  {Kim}, {Haan}, {Inami}, {Murphy}, {Appleton}, {Barnes}, {Bothun}, {Bridge},
  {Charmandaris}, {Jensen}, {Kewley}, {Lord}, {Madore}, {Marshall},
  {Melbourne}, {Rich}, {Satyapal}, {Schulz}, {Spoon}, {Sturm}, {U}, {Veilleux},
  \& {Xu}}]{Armus:2009aa}
{Armus}, L., {Mazzarella}, J.~M., {Evans}, A.~S., {et~al.} 2009, \pasp, 121,
  559

\bibitem[{{Baan} \& {Haschick}(1987)}]{Baan:1987aa}
{Baan}, W.~A., \& {Haschick}, A.~D. 1987, \apj, 318, 139

\bibitem[{{Barnes}(2001)}]{Barnes:2001aa}
{Barnes}, J.~E. 2001, in Astronomical Society of the Pacific Conference Series,
  Vol. 245, Astrophysical Ages and Times Scales, ed. T.~{von Hippel},
  C.~{Simpson}, \& N.~{Manset}, 382

\bibitem[{{Barnes}(2002)}]{Barnes:2002aa}
{Barnes}, J.~E. 2002, \mnras, 333, 481

\bibitem[{{Barnes}(2004)}]{Barnes:2004aa}
---. 2004, \mnras, 350, 798

\bibitem[{{Barnes} \& {Hernquist}(1992)}]{Barnes:1992aa}
{Barnes}, J.~E., \& {Hernquist}, L. 1992, \nat, 360, 715

\bibitem[{{Barnes} \& {Hernquist}(1996)}]{Barnes:1996aa}
---. 1996, \apj, 471, 115

\bibitem[{{Barnes} \& {Hibbard}(2009)}]{Barnes:2009aa}
{Barnes}, J.~E., \& {Hibbard}, J.~E. 2009, \aj, 137, 3071

\bibitem[{{Bolatto} {et~al.}(2013){Bolatto}, {Wolfire}, \&
  {Leroy}}]{Bolatto:2013aa}
{Bolatto}, A.~D., {Wolfire}, M., \& {Leroy}, A.~K. 2013, \araa, 51, 207

\bibitem[{{Chien}(2010)}]{Chien:2010aa}
{Chien}, L. 2010, in Astronomical Society of the Pacific Conference Series,
  Vol. 423, Galaxy Wars: Stellar Populations and Star Formation in Interacting
  Galaxies, ed. B.~{Smith}, J.~{Higdon}, S.~{Higdon}, \& N.~{Bastian}, 197

\bibitem[{{Downes} \& {Solomon}(1998)}]{Downes:1998aa}
{Downes}, D., \& {Solomon}, P.~M. 1998, \apj, 507, 615

\bibitem[{{Draine} {et~al.}(2007){Draine}, {Dale}, {Bendo}, {Gordon}, {Smith},
  {Armus}, {Engelbracht}, {Helou}, {Kennicutt}, {Li}, {Roussel}, {Walter},
  {Calzetti}, {Moustakas}, {Murphy}, {Rieke}, {Bot}, {Hollenbach}, {Sheth}, \&
  {Teplitz}}]{Draine:2007aa}
{Draine}, B.~T., {Dale}, D.~A., {Bendo}, G., {et~al.} 2007, \apj, 663, 866

\bibitem[{{Elbaz} {et~al.}(2007){Elbaz}, {Daddi}, {Le Borgne}, {Dickinson},
  {Alexander}, {Chary}, {Starck}, {Brandt}, {Kitzbichler}, {MacDonald},
  {Nonino}, {Popesso}, {Stern}, \& {Vanzella}}]{Elbaz:2007aa}
{Elbaz}, D., {Daddi}, E., {Le Borgne}, D., {et~al.} 2007, \aap, 468, 33

\bibitem[{{Fischer} {et~al.}(2010){Fischer}, {Sturm}, {Gonz{\'a}lez-Alfonso},
  {Graci{\'a}-Carpio}, {Hailey-Dunsheath}, {Poglitsch}, {Contursi}, {Lutz},
  {Genzel}, {Sternberg}, {Verma}, \& {Tacconi}}]{Fischer:2010aa}
{Fischer}, J., {Sturm}, E., {Gonz{\'a}lez-Alfonso}, E., {et~al.} 2010, \aap,
  518, L41

\bibitem[{{Gao} {et~al.}(2001){Gao}, {Lo}, {Lee}, \& {Lee}}]{Gao:2001aa}
{Gao}, Y., {Lo}, K.~Y., {Lee}, S.-W., \& {Lee}, T.-H. 2001, \apj, 548, 172

\bibitem[{{Genzel} {et~al.}(2001){Genzel}, {Tacconi}, {Rigopoulou}, {Lutz}, \&
  {Tecza}}]{Genzel:2001aa}
{Genzel}, R., {Tacconi}, L.~J., {Rigopoulou}, D., {Lutz}, D., \& {Tecza}, M.
  2001, \apj, 563, 527

\bibitem[{{Genzel} {et~al.}(2014){Genzel}, {F{\"o}rster Schreiber}, {Rosario},
  {Lang}, {Lutz}, {Wisnioski}, {Wuyts}, {Wuyts}, {Bandara}, {Bender}, {Berta},
  {Kurk}, {Mendel}, {Tacconi}, {Wilman}, {Beifiori}, {Brammer}, {Burkert},
  {Buschkamp}, {Chan}, {Carollo}, {Davies}, {Eisenhauer}, {Fabricius},
  {Fossati}, {Kriek}, {Kulkarni}, {Lilly}, {Mancini}, {Momcheva}, {Naab},
  {Nelson}, {Renzini}, {Saglia}, {Sharples}, {Sternberg}, {Tacchella}, \& {van
  Dokkum}}]{Genzel:2014aa}
{Genzel}, R., {F{\"o}rster Schreiber}, N.~M., {Rosario}, D., {et~al.} 2014,
  \apj, 796, 7

\bibitem[{{Graham} {et~al.}(1990){Graham}, {Carico}, {Matthews}, {Neugebauer},
  {Soifer}, \& {Wilson}}]{Graham:1990aa}
{Graham}, J.~R., {Carico}, D.~P., {Matthews}, K., {et~al.} 1990, \apjl, 354, L5

\bibitem[{{Haan} {et~al.}(2011){Haan}, {Surace}, {Armus}, {Evans}, {Howell},
  {Mazzarella}, {Kim}, {Vavilkin}, {Inami}, {Sanders}, {Petric}, {Bridge},
  {Melbourne}, {Charmandaris}, {Diaz-Santos}, {Murphy}, {U}, {Stierwalt}, \&
  {Marshall}}]{Haan:2011aa}
{Haan}, S., {Surace}, J.~A., {Armus}, L., {et~al.} 2011, \aj, 141, 100

\bibitem[{{Hopkins} {et~al.}(2013){Hopkins}, {Cox}, {Hernquist}, {Narayanan},
  {Hayward}, \& {Murray}}]{Hopkins:2013aa}
{Hopkins}, P.~F., {Cox}, T.~J., {Hernquist}, L., {et~al.} 2013, \mnras, 430,
  1901

\bibitem[{{Hung} {et~al.}(2013){Hung}, {Sanders}, {Casey}, {Lee}, {Barnes},
  {Capak}, {Kartaltepe}, {Koss}, {Larson}, {Le Floc'h}, {Lockhart}, {Man},
  {Mann}, {Riguccini}, {Scoville}, \& {Symeonidis}}]{Hung:2013aa}
{Hung}, C.-L., {Sanders}, D.~B., {Casey}, C.~M., {et~al.} 2013, \apj, 778, 129

\bibitem[{{Ishida}(2004)}]{Ishida:2004aa}
{Ishida}, C.~M. 2004, PhD thesis, UNIVERSITY OF HAWAI'I

\bibitem[{{Iwasawa} {et~al.}(2011){Iwasawa}, {Sanders}, {Teng}, {U}, {Armus},
  {Evans}, {Howell}, {Komossa}, {Mazzarella}, {Petric}, {Surace}, {Vavilkin},
  {Veilleux}, \& {Trentham}}]{Iwasawa:2011aa}
{Iwasawa}, K., {Sanders}, D.~B., {Teng}, S.~H., {et~al.} 2011, \aap, 529, A106

\bibitem[{{Joseph}(1999)}]{Joseph:1999aa}
{Joseph}, R.~D. 1999, \apss, 266, 321

\bibitem[{{Joseph} \& {Wright}(1985)}]{Joseph:1985aa}
{Joseph}, R.~D., \& {Wright}, G.~S. 1985, \mnras, 214, 87

\bibitem[{{Kewley} {et~al.}(2010){Kewley}, {Rupke}, {Zahid}, {Geller}, \&
  {Barton}}]{Kewley:2010aa}
{Kewley}, L.~J., {Rupke}, D., {Zahid}, H.~J., {Geller}, M.~J., \& {Barton},
  E.~J. 2010, \apjl, 721, L48

\bibitem[{{Kim} \& {Sanders}(1998)}]{Kim:1998ab}
{Kim}, D.-C., \& {Sanders}, D.~B. 1998, \apjs, 119, 41

\bibitem[{{Kim} {et~al.}(1998){Kim}, {Veilleux}, \& {Sanders}}]{Kim:1998aa}
{Kim}, D.-C., {Veilleux}, S., \& {Sanders}, D.~B. 1998, \apj, 508, 627

\bibitem[{{Kim} {et~al.}(2013){Kim}, {Evans}, {Vavilkin}, {Armus},
  {Mazzarella}, {Sheth}, {Surace}, {Haan}, {Howell}, {D{\'{\i}}az-Santos},
  {Petric}, {Iwasawa}, {Privon}, \& {Sanders}}]{Kim:2013aa}
{Kim}, D.-C., {Evans}, A.~S., {Vavilkin}, T., {et~al.} 2013, \apj, 768, 102

\bibitem[{{Komatsu} {et~al.}(2009){Komatsu}, {Dunkley}, {Nolta}, {Bennett},
  {Gold}, {Hinshaw}, {Jarosik}, {Larson}, {Limon}, {Page}, {Spergel},
  {Halpern}, {Hill}, {Kogut}, {Meyer}, {Tucker}, {Weiland}, {Wollack}, \&
  {Wright}}]{Komatsu:2009aa}
{Komatsu}, E., {Dunkley}, J., {Nolta}, M.~R., {et~al.} 2009, \apjs, 180, 330

\bibitem[{{Le Floc'h} {et~al.}(2005){Le Floc'h}, {Papovich}, {Dole}, {Bell},
  {Lagache}, {Rieke}, {Egami}, {P{\'e}rez-Gonz{\'a}lez}, {Alonso-Herrero},
  {Rieke}, {Blaylock}, {Engelbracht}, {Gordon}, {Hines}, {Misselt}, {Morrison},
  \& {Mould}}]{LeFloch:2005aa}
{Le Floc'h}, E., {Papovich}, C., {Dole}, H., {et~al.} 2005, \apj, 632, 169

\bibitem[{{Leitherer} {et~al.}(1999){Leitherer}, {Schaerer}, {Goldader},
  {Delgado}, {Robert}, {Kune}, {de Mello}, {Devost}, \&
  {Heckman}}]{Leitherer:1999aa}
{Leitherer}, C., {Schaerer}, D., {Goldader}, J.~D., {et~al.} 1999, \apjs, 123,
  3

\bibitem[{{Magnier} {et~al.}(2013){Magnier}, {Schlafly}, {Finkbeiner}, {Juric},
  {Tonry}, {Burgett}, {Chambers}, {Flewelling}, {Kaiser}, {Kudritzki},
  {Morgan}, {Price}, {Sweeney}, \& {Stubbs}}]{Magnier:2013aa}
{Magnier}, E.~A., {Schlafly}, E., {Finkbeiner}, D., {et~al.} 2013, \apjs, 205,
  20

\bibitem[{{Mazzarella} {et~al.}(2012){Mazzarella}, {Iwasawa}, {Vavilkin},
  {Armus}, {Kim}, {Bothun}, {Evans}, {Spoon}, {Haan}, {Howell}, {Lord},
  {Marshall}, {Ishida}, {Xu}, {Petric}, {Sanders}, {Surace}, {Appleton},
  {Chan}, {Frayer}, {Inami}, {Khachikian}, {Madore}, {Privon}, {Sturm}, {U}, \&
  {Veilleux}}]{Mazzarella:2012aa}
{Mazzarella}, J.~M., {Iwasawa}, K., {Vavilkin}, T., {et~al.} 2012, \aj, 144,
  125

\bibitem[{{Mihos} \& {Hernquist}(1996)}]{Mihos:1996aa}
{Mihos}, J.~C., \& {Hernquist}, L. 1996, \apj, 464, 641

\bibitem[{{Mirabel} {et~al.}(1990){Mirabel}, {Booth}, {Johansson}, {Garay}, \&
  {Sanders}}]{Mirabel:1990aa}
{Mirabel}, I.~F., {Booth}, R.~S., {Johansson}, L.~E.~B., {Garay}, G., \&
  {Sanders}, D.~B. 1990, \aap, 236, 327

\bibitem[{{Mirabel} {et~al.}(1988){Mirabel}, {Kazes}, \&
  {Sanders}}]{Mirabel:1988aa}
{Mirabel}, I.~F., {Kazes}, I., \& {Sanders}, D.~B. 1988, \apjl, 324, L59

\bibitem[{{Mirabel} \& {Sanders}(1989)}]{Mirabel:1989aa}
{Mirabel}, I.~F., \& {Sanders}, D.~B. 1989, \apjl, 340, L53

\bibitem[{{Privon} {et~al.}(2013){Privon}, {Barnes}, {Evans}, {Hibbard}, {Yun},
  {Mazzarella}, {Armus}, \& {Surace}}]{Privon:2013aa}
{Privon}, G.~C., {Barnes}, J.~E., {Evans}, A.~S., {et~al.} 2013, \apj, 771, 120

\bibitem[{{Rich} {et~al.}(2012){Rich}, {Torrey}, {Kewley}, {Dopita}, \&
  {Rupke}}]{Rich:2012aa}
{Rich}, J.~A., {Torrey}, P., {Kewley}, L.~J., {Dopita}, M.~A., \& {Rupke},
  D.~S.~N. 2012, \apj, 753, 5

\bibitem[{{Rieke} \& {Low}(1972)}]{Rieke:1972aa}
{Rieke}, G.~H., \& {Low}, F.~J. 1972, \apjl, 176, L95

\bibitem[{{Rothberg} \& {Joseph}(2006)}]{Rothberg:2006aa}
{Rothberg}, B., \& {Joseph}, R.~D. 2006, \aj, 132, 976

\bibitem[{{Sanders}(1999)}]{Sanders:1999aa}
{Sanders}, D.~B. 1999, \apss, 266, 331

\bibitem[{{Sanders} {et~al.}(2003){Sanders}, {Mazzarella}, {Kim}, {Surace}, \&
  {Soifer}}]{Sanders:2003aa}
{Sanders}, D.~B., {Mazzarella}, J.~M., {Kim}, D.-C., {Surace}, J.~A., \&
  {Soifer}, B.~T. 2003, \aj, 126, 1607

\bibitem[{{Sanders} \& {Mirabel}(1985)}]{Sanders:1985aa}
{Sanders}, D.~B., \& {Mirabel}, I.~F. 1985, \apjl, 298, L31

\bibitem[{{Sanders} {et~al.}(1991){Sanders}, {Scoville}, \&
  {Soifer}}]{Sanders:1991aa}
{Sanders}, D.~B., {Scoville}, N.~Z., \& {Soifer}, B.~T. 1991, \apj, 370, 158

\bibitem[{{Sanders} {et~al.}(1986){Sanders}, {Scoville}, {Young}, {Soifer},
  {Schloerb}, {Rice}, \& {Danielson}}]{Sanders:1986aa}
{Sanders}, D.~B., {Scoville}, N.~Z., {Young}, J.~S., {et~al.} 1986, \apjl, 305,
  L45

\bibitem[{{Sanders} {et~al.}(1989){Sanders}, {Scoville}, {Zensus}, {Soifer},
  {Wilson}, {Zylka}, \& {Steppe}}]{Sanders:1989aa}
{Sanders}, D.~B., {Scoville}, N.~Z., {Zensus}, A., {et~al.} 1989, \aap, 213, L5

\bibitem[{{Sanders} {et~al.}(1988{\natexlab{a}}){Sanders}, {Soifer}, {Elias},
  {Madore}, {Matthews}, {Neugebauer}, \& {Scoville}}]{Sanders:1988aa}
{Sanders}, D.~B., {Soifer}, B.~T., {Elias}, J.~H., {et~al.} 1988{\natexlab{a}},
  \apj, 325, 74

\bibitem[{{Sanders} {et~al.}(1988{\natexlab{b}}){Sanders}, {Soifer}, {Elias},
  {Neugebauer}, \& {Matthews}}]{Sanders:1988ab}
{Sanders}, D.~B., {Soifer}, B.~T., {Elias}, J.~H., {Neugebauer}, G., \&
  {Matthews}, K. 1988{\natexlab{b}}, \apjl, 328, L35

\bibitem[{{Schlafly} {et~al.}(2012){Schlafly}, {Finkbeiner}, {Juri{\'c}},
  {Magnier}, {Burgett}, {Chambers}, {Grav}, {Hodapp}, {Kaiser}, {Kudritzki},
  {Martin}, {Morgan}, {Price}, {Rix}, {Stubbs}, {Tonry}, \&
  {Wainscoat}}]{Schlafly:2012aa}
{Schlafly}, E.~F., {Finkbeiner}, D.~P., {Juri{\'c}}, M., {et~al.} 2012, \apj,
  756, 158

\bibitem[{{Schweizer} \& {Seitzer}(1992)}]{Schweizer:1992aa}
{Schweizer}, F., \& {Seitzer}, P. 1992, \aj, 104, 1039

\bibitem[{{Schweizer} \& {Seitzer}(2007)}]{Schweizer:2007aa}
---. 2007, \aj, 133, 2132

\bibitem[{{Scoville} {et~al.}(2016){Scoville}, {Sheth}, {Aussel}, {Vanden
  Bout}, {Capak}, {Bongiorno}, {Casey}, {Murchikova}, {Koda},
  {{\'A}lvarez-M{\'a}rquez}, {Lee}, {Laigle}, {McCracken}, {Ilbert}, {Pope},
  {Sanders}, {Chu}, {Toft}, {Ivison}, \& {Manohar}}]{Scoville:2016aa}
{Scoville}, N., {Sheth}, K., {Aussel}, H., {et~al.} 2016, \apj, 820, 83

\bibitem[{{Scoville} {et~al.}(1987){Scoville}, {Yun}, {Sanders}, {Clemens}, \&
  {Waller}}]{Scoville:1987aa}
{Scoville}, N.~Z., {Yun}, M.~S., {Sanders}, D.~B., {Clemens}, D.~P., \&
  {Waller}, W.~H. 1987, \apjs, 63, 821

\bibitem[{{Soifer} {et~al.}(1987){Soifer}, {Sanders}, {Madore}, {Neugebauer},
  {Danielson}, {Elias}, {Lonsdale}, \& {Rice}}]{Soifer:1987aa}
{Soifer}, B.~T., {Sanders}, D.~B., {Madore}, B.~F., {et~al.} 1987, \apj, 320,
  238

\bibitem[{{Solomon} {et~al.}(1997){Solomon}, {Downes}, {Radford}, \&
  {Barrett}}]{Solomon:1997aa}
{Solomon}, P.~M., {Downes}, D., {Radford}, S.~J.~E., \& {Barrett}, J.~W. 1997,
  \apj, 478, 144

\bibitem[{{Solomon} {et~al.}(1987){Solomon}, {Rivolo}, {Barrett}, \&
  {Yahil}}]{Solomon:1987aa}
{Solomon}, P.~M., {Rivolo}, A.~R., {Barrett}, J., \& {Yahil}, A. 1987, \apj,
  319, 730

\bibitem[{{Stierwalt} {et~al.}(2013){Stierwalt}, {Armus}, {Surace}, {Inami},
  {Petric}, {Diaz-Santos}, {Haan}, {Charmandaris}, {Howell}, {Kim}, {Marshall},
  {Mazzarella}, {Spoon}, {Veilleux}, {Evans}, {Sanders}, {Appleton}, {Bothun},
  {Bridge}, {Chan}, {Frayer}, {Iwasawa}, {Kewley}, {Lord}, {Madore},
  {Melbourne}, {Murphy}, {Rich}, {Schulz}, {Sturm}, {Vavilkin}, \&
  {Xu}}]{Stierwalt:2013aa}
{Stierwalt}, S., {Armus}, L., {Surace}, J.~A., {et~al.} 2013, \apjs, 206, 1

\bibitem[{{Sturm} {et~al.}(2011){Sturm}, {Gonz{\'a}lez-Alfonso}, {Veilleux},
  {Fischer}, {Graci{\'a}-Carpio}, {Hailey-Dunsheath}, {Contursi}, {Poglitsch},
  {Sternberg}, {Davies}, {Genzel}, {Lutz}, {Tacconi}, {Verma}, {Maiolino}, \&
  {de Jong}}]{Sturm:2011aa}
{Sturm}, E., {Gonz{\'a}lez-Alfonso}, E., {Veilleux}, S., {et~al.} 2011, \apjl,
  733, L16

\bibitem[{{Surace} {et~al.}(1998){Surace}, {Sanders}, {Vacca}, {Veilleux}, \&
  {Mazzarella}}]{Surace:1998aa}
{Surace}, J.~A., {Sanders}, D.~B., {Vacca}, W.~D., {Veilleux}, S., \&
  {Mazzarella}, J.~M. 1998, \apj, 492, 116

\bibitem[{{Tacconi} {et~al.}(2008){Tacconi}, {Genzel}, {Smail}, {Neri},
  {Chapman}, {Ivison}, {Blain}, {Cox}, {Omont}, {Bertoldi}, {Greve},
  {F{\"o}rster Schreiber}, {Genel}, {Lutz}, {Swinbank}, {Shapley}, {Erb},
  {Cimatti}, {Daddi}, \& {Baker}}]{Tacconi:2008aa}
{Tacconi}, L.~J., {Genzel}, R., {Smail}, I., {et~al.} 2008, \apj, 680, 246

\bibitem[{{Tombesi} {et~al.}(2015){Tombesi}, {Mel{\'e}ndez}, {Veilleux},
  {Reeves}, {Gonz{\'a}lez-Alfonso}, \& {Reynolds}}]{Tombesi:2015aa}
{Tombesi}, F., {Mel{\'e}ndez}, M., {Veilleux}, S., {et~al.} 2015, \nat, 519,
  436

\bibitem[{{Tonry} {et~al.}(2012){Tonry}, {Stubbs}, {Lykke}, {Doherty},
  {Shivvers}, {Burgett}, {Chambers}, {Hodapp}, {Kaiser}, {Kudritzki},
  {Magnier}, {Morgan}, {Price}, \& {Wainscoat}}]{Tonry:2012aa}
{Tonry}, J.~L., {Stubbs}, C.~W., {Lykke}, K.~R., {et~al.} 2012, \apj, 750, 99

\bibitem[{{Toomre} \& {Toomre}(1972)}]{Toomre:1972aa}
{Toomre}, A., \& {Toomre}, J. 1972, \apj, 178, 623

\bibitem[{{U} {et~al.}(2012){U}, {Sanders}, {Mazzarella}, {Evans}, {Howell},
  {Surace}, {Armus}, {Iwasawa}, {Kim}, {Casey}, {Vavilkin}, {Dufault},
  {Larson}, {Barnes}, {Chan}, {Frayer}, {Haan}, {Inami}, {Ishida},
  {Kartaltepe}, {Melbourne}, \& {Petric}}]{U:2012aa}
{U}, V., {Sanders}, D.~B., {Mazzarella}, J.~M., {et~al.} 2012, \apjs, 203, 9

\bibitem[{{U} {et~al.}(2013){U}, {Medling}, {Sanders}, {Max}, {Armus},
  {Iwasawa}, {Evans}, {Kewley}, \& {Fazio}}]{U:2013aa}
{U}, V., {Medling}, A., {Sanders}, D., {et~al.} 2013, \apj, 775, 115

\bibitem[{{V{\"a}is{\"a}nen} {et~al.}(2012){V{\"a}is{\"a}nen}, {Rajpaul},
  {Zijlstra}, {Reunanen}, \& {Kotilainen}}]{Vaisanen:2012aa}
{V{\"a}is{\"a}nen}, P., {Rajpaul}, V., {Zijlstra}, A.~A., {Reunanen}, J., \&
  {Kotilainen}, J. 2012, \mnras, 420, 2209

\bibitem[{{Veilleux} {et~al.}(2013){Veilleux}, {Mel{\'e}ndez}, {Sturm},
  {Gracia-Carpio}, {Fischer}, {Gonz{\'a}lez-Alfonso}, {Contursi}, {Lutz},
  {Poglitsch}, {Davies}, {Genzel}, {Tacconi}, {de Jong}, {Sternberg}, {Netzer},
  {Hailey-Dunsheath}, {Verma}, {Rupke}, {Maiolino}, {Teng}, \&
  {Polisensky}}]{Veilleux:2013aa}
{Veilleux}, S., {Mel{\'e}ndez}, M., {Sturm}, E., {et~al.} 2013, \apj, 776, 27

\bibitem[{{White} \& {Rees}(1978)}]{White:1978aa}
{White}, S.~D.~M., \& {Rees}, M.~J. 1978, \mnras, 183, 341

\bibitem[{{Xu} {et~al.}(2015){Xu}, {Cao}, {Lu}, {Gao}, {Diaz-Santos},
  {Herrero-Illana}, {Meijerink}, {Privon}, {Zhao}, {Evans}, {K{\"o}nig},
  {Mazzarella}, {Aalto}, {Appleton}, {Armus}, {Charmandaris}, {Chu}, {Haan},
  {Inami}, {Murphy}, {Sanders}, {Schulz}, \& {van der Werf}}]{Xu:2015aa}
{Xu}, C.~K., {Cao}, C., {Lu}, N., {et~al.} 2015, \apj, 799, 11

\bibitem[{{Yuan} {et~al.}(2010){Yuan}, {Kewley}, \& {Sanders}}]{Yuan:2010aa}
{Yuan}, T.-T., {Kewley}, L.~J., \& {Sanders}, D.~B. 2010, \apj, 709, 884

\bibitem[{{Zhu} {et~al.}(2003){Zhu}, {Seaquist}, \& {Kuno}}]{Zhu:2003aa}
{Zhu}, M., {Seaquist}, E.~R., \& {Kuno}, N. 2003, \apj, 588, 243

\end{thebibliography}

 \end{document}